\documentclass[preprint]{aastex}
\usepackage{multirow}

\def\kms{{\rm\,km\,s^{-1}}}
\def\kmskpc{{\rm\,km\, \,s^{-1} \, {kpc}^{-1}}}

\def\deg{{^\circ}}

\def\kpc{{\rm\,kpc}}

\def\mathnew{\mathsurround=0pt}   
\def\simov#1#2{\lower .5pt\vbox{\baselineskip0pt  
    \lineskip-.5pt\ialign{$\mathnew#1\hfil##\hfil$\crcr#2\crcr\sim\crcr}}}    
    
\def\simless{\mathrel{\mathpalette\simov <}}   
\def\'#1{\ifx#1i{\accent"13\i}\else{\accent"13#1}\fi}    
  
\def\et{et~al. }     
 
\received{}
\begin{document}

\shorttitle{Restrictions for Spiral Arms in Normal Spiral Galaxies}  
\shortauthors{P\'erez-Villegas et al. 2015}

\title{Stellar Orbital Studies in Normal Spiral Galaxies II: Restrictions to Structural and Dynamical parameters on Spiral Arms}

\author{A. P\'erez-Villegas$^{1}$, B. Pichardo$^{2}$, E. Moreno$^{2}$}     

\affil{$^{1}$ Max-Planck-Instit\"ut f\"ur Extraterrestrische Physik, Gie\ss enbachstra\ss e, 85748 Garching, Germany; \\mperez@mpe.mpg.de}

\affil{$^{2}$Instituto de Astronom\'ia, Universidad Nacional
Aut\'onoma de M\'exico, A.P. 70--264, 04510, M\'exico, D.F.; Universitaria, D.F., M\'exico}

\begin{abstract}
Making use of a set of detailed potential models for normal spiral galaxies, we analyze the disk stellar orbital dynamics as the structural and dynamical parameters of the spiral arms (mass, pattern speed and pitch angle) are gradually modified. With this comprehensive study of ordered and chaotic behavior, we constructed an assemblage of orbitally supported galactic models and plausible parameters for orbitally self-consistent spiral arms models. We find that, to maintain orbital support for the spiral arms, the spiral arm mass, M$_{sp}$, must decrease with the increase of the pitch angle, $i$; if $i$ is smaller than $\sim10\deg$, M$_{sp}$ can be as large as $\sim7\%$, $\sim6\%$, $\sim5\%$ of the disk mass, for Sa, Sb, and Sc galaxies, respectively. If $i$ increases up to $\sim25\deg$, the maximum M$_{sp}$ is $\sim1\%$ of the disk mass independently in this case of morphological type. For values larger than these limits, spiral arms would likely act as transient features. Regarding the limits posed by extreme chaotic behavior, we find a strong restriction on the maximum plausible values of spiral arms parameters on disk galaxies beyond which, chaotic behavior becomes pervasive. We find that for $i$ smaller than $\sim20\deg$, $\sim25\deg$, $\sim30\deg$, for Sa, Sb, and Sc galaxies, respectively, M$_{sp}$ can go up to $\sim10\%$, of the mass of the disk. If the corresponding $i$ is around $\sim40\deg$, $\sim45\deg$, $\sim50\deg$, M$_{sp}$ is $\sim1\%$, $\sim2\%$, $\sim3\%$ of the mass of the disk. Beyond these values, chaos dominates phase space, destroying the main periodic and the neighboring quasi-periodic orbits.

\end{abstract}

\keywords{Chaos -- galaxies: evolution -- galaxies: kinematics and
dynamics -- galaxies: spiral -- galaxies: structure}

\section{Introduction} \label{sec:intro}    

With the advent of the new extended and profound surveys of the Galaxy
and other galaxies, we will likely understand much more of spiral
galaxy morphology and kinematics with unprecedented detail. At this
moment however, our understanding of splendid structures in galaxies,
such as the spiral arms, is quite limited; including for example, its
very nature and origin, how they are supported, whether they are long
lasting or transient, if they exhibit noticeable (observable) effects
on the stellar and gaseous dynamics behavior, what are their orbital
effects in different types of spirals, etc.

The first firm step into the path of understanding the Milky Way and
spiral galaxies in general was given by morphological classifications,
that have provided important statistical information about their
structural parameters, such as luminosity ratios of their main
components (bulge, disk, and nonaxisymmetric large scale features:
bars, spiral arms, rings, etc.), rotation curve, spiral-arms pitch
angles, etc. The first morphological classification that tried to
taxonomize galaxies was the Hubble sequence (Hubble 1926, 1936).  This
represents the simplest classification scheme and within it, the
normal spiral galaxies, in which we focus this work, range from
`early' to `late' (Sa to Sc), mainly based on two criteria: the pitch
angle of spiral arms and the bulge-to-disk luminosity ratio.  An Sa
galaxy possesses smoother closed spiral arms and a conspicuous central
bulge, an Sc galaxy has open and remarkably structured spiral
arms and a small central bulge, an Sb galaxy is intermediate between
both types. Although the Hubble morphological classification is
satisfactory for galaxies with redshift $z < 0.5$ (van den Bergh
2002), where non-interacting galaxies have had time to relax, it is
well known that this scheme does not fully comprises all
galaxies. Indeed, structural parameters of disks present a large
scatter among catalogued galaxies with the same morphological
type. The best example of this is the pitch angle, that ranges from
about $8\deg$ to $50\deg$ (Kennicutt 1981; Ma \et 2000; Davis \et
2012) for late type galaxies, for instance.

Some recent studies present spiral arms as likely transient features
from simulations (D'Onghia \et 2013; Baba \et 2013; Grand \et 2012a,b;
Ro{\v s}kar \et 2012; Wada \et 2011; Sellwood 2011; Fujii \et 2011;
Dobbs \& Bonnell 2006), or transient as a product of overlapping of
multiple spiral modes coupled together through resonances at the
corotation radius (Sellwood \& Carlberg 2014; Ro{\v s}kar \et 2012;
Quillen \et 2011). Spiral arms have also been found to corotate
  with the disk (i.e. winding up, therefore transient ; Roca-F\`abrega
  \et 2013; Kawata \et 2014). Also, from the observational point of
view there seem to be evidences of transient spiral arms (Speights \&
Westpfahl 2011, 2012; Ferreras \et 2012; Foyle \et 2011; Meidt \et
2008; Meidt \et 2009; Merrifield \et 2006). On the other hand,
observational and theoretical evidences of the opposite, i.e.,
long-lived spiral arms, has been presented (Scarano \& L\'epine 2013;
Cedr\'es \et 2013; Mart\'inez-Garc\'ia \& Gonz\'alez-Lopezlira 2013;
Law \et 2012; Scarano \et 2011; S\'anchez-Gil \et 2011; Egusa \et
2009; Grosb\o l \& Dottori 2009; Zhang 1998; Donner \& Thomasson 1994;
Efremov 1985). Whether spiral arms are all transient features, or in
some cases they could be long-lasting, remains still a polemic matter
in modern astrophysics.

In previous studies (P\'erez-Villegas \et 2013, hereafter Paper I, and
P\'erez-Villegas \et 2012), we employed the ideal Hubble
classification scheme as the base to construct the axisymmetric
background gravitational potential models for spiral galaxies
(i.e. bulge, disk and halo). With this set of models, we analyzed the
stellar orbital dynamics on disks, produced by spiral arms in
different galactic morphological types. The constructed galactic
potential models then follow the typical morphology that characterizes
the Hubble sequence in terms of the rotation curve, bulge-to-disk mass
ratios and scale-lengths. These studies were performed on steady,
realistic potentials for spiral galaxies. These type of potentials can
not follow the galactic evolution, but they are able to provide some
restrictions of potentials based on the detailed structure of orbital
chaos, and on the existence and structure of periodic orbits as the
dynamical support to shape stellar systems. The main purpose of our
study has been to disentangle all possible details of the orbital
structure, which are not straightforward to discern yet on N-body
simulations.

Using these models, we performed an extensive study of the pitch angle
in normal spiral galaxies. We run thousands of orbits for different
timescales depending on the specific problem. Two restrictions to the
spiral arms structure were imposed theoretically; one on their steady
or transient nature and the other on their maximum pitch angle prior
to destruction. The first restriction is based on the orbital ordered
behavior, where we found a maximum pitch angle of $\sim15\deg$,
$\sim18\deg$ and $\sim20\deg$ for Sa, Sb and Sc spiral galaxies,
respectively. Up to these limits the density response supports closely
the imposed spiral arms at all radii, the spiral arms are stable, and
could be of long-lasting nature. Galaxies with spiral arms having
pitch angles beyond these limits would rather be explained as
transient features.  The second restriction is based on chaotic
orbital behavior; in this case the limits for the pitch angle are
$\sim30\deg$, $\sim40\deg$ and $\sim50\deg$ for Sa, Sb and Sc spiral
galaxies, respectively. Beyond these limits, chaos becomes pervasive
wiping out the spiral arms.

In the present analysis we continue our extensive stellar dynamical
studies in normal spiral galaxies, from early to late types, but now
we cover all the most important spiral arms parameters that include:
arms total mass (relative to the disk), angular velocity, and pitch
angle interrelated. We produce experiments exploring the statistical
effect on stellar orbits on the galactic plane due to the variations
of these parameters. The main objectives of this work are (1) to
elucidate the influence of spiral arms on different morphological
types of galaxies, as going from early to late types, (2) to provide
some restrictions to structural and dynamical parameters of galaxies,
and (3) to produce a set of parameters for `allowed spiral models',
which are self-consistent from an orbital (periodic orbits) point of
view, with good probabilities of being long-lasting structures, and
with mild or quiet chaotic nature. With these parameters, steady
models can be constructed that result in likely robust and persisting
entities.

This paper is organized as follows. The galactic potential and the
methodology are described in Section \ref{model}. The effect on the
disk dynamics, due to the variation of the spral arms mass and their
angular velocity in different morphological types (Sa, Sb and Sc
galaxies) is presented in Section \ref{results}. In Section
\ref{valid_parameters}, we present a valid set of structural and
dynamical parameters for plausible long-lasting spiral arms nature,
and also their maximum values before chaos dominates. In Section
\ref{discussion} we discuss the effect of structural and dynamical
parameters of spiral arms (pitch angle, angular speed, and mass) in
normal spiral galaxies and present our conclusions.

\section{Numerical Implementation and Methodology}\label{model}

With the use of the observationally motivated family of models for
normal spiral galactic potentials presented in Paper I, we performed a
comprehensive stellar orbital dynamics study. The main tools employed
for this task are periodic orbits, density response calculations, and
phase-space (Poincar\'e) diagrams. The potential of each galaxy is
formed by an axisymmetric part, plus a nonaxisymmetric potential
represented by a detailed model of the spiral arms. In the following,
we summarize some properties of the galactic models and the employed
tools.

\subsection{Models for Normal Spiral Galaxies}\label{normal_spiral}

The galactic models consist of axisymmetric and nonaxisymmetric parts.
The axisymmetric part is formed by a bulge and disk of the form
proposed by Miyamoto-Nagai (1975), and a massive spherical halo
(Allen \& Santill\'an 1991). With these components, in
Paper I we fit the different galaxy types considering the typical
rotation curves for Sa, Sb and Sc galaxies, and the bulge-to-disk mass
ratios (see Figure 1 of Paper I). The nonaxisymmetric part is the
three-dimensional model of spiral arms given by Pichardo \et 2003,
called PERLAS. This model is a mass distribution formed by a set of 
inhomogeneous oblate spheroids lying on a logarithmic spiral locus; it
has been tested and compared with other theoretical models 
(Martos et al. 2004; Antoja et al. 2009, 2011). 

In Table \ref{tab:parameters} we present the observational and
theoretical parameters employed to fit the galactic models and their
respective references (data taken from Paper I). In this table,
D, B, H, refer to the disk, bulge, and halo components, respectively;
M$_{\rm sp}$ is the total mass of the spiral arms, $i$ is their pitch
angle, and $\Omega_{\rm p}$ is their angular speed. In our models we
take a clockwise rotation. To simplify the notation, in the following
we call $\mu$ = M$_{\rm sp}$/M$_{\rm D}$, i.e., the ratio of the mass
of the spiral arms to the mass of the disk.

 \begin{deluxetable}{lcccl}
\tablecolumns{5}
\tabletypesize{\scriptsize}
\tablewidth{0pt}
\tablecaption{Parameters of the Axisymmetric Potential}
\tablehead{{Parameter} &\multicolumn{3}{c}{Value}& {Reference}}
\startdata

 &\multicolumn {3}{c}{Spiral Arms} \\
\hline
  &Sa&Sb&Sc &\\
\hline
Locus             & \multicolumn{3}{c}{Logarithmic } & 1,9,10\\
Arms Number       & \multicolumn{3}{c}{2} & 2\\
Pitch Angle $i$ ($\deg$)       & 7 -- 20& 10 -- 20& 15 -- 30& 3,7 \\
$\mu$ = M$_{\rm sp}$/M$_{\rm D}$& \multicolumn{3}{c}{0.01 -- 0.10} &  \\
Scale-Length ($\kpc$)    & 7&5& 3  & Disk based\\
$\Omega_{\rm p}$ $^1$ ($\kmskpc$) &\multicolumn{3}{c} {$10$ to $60$}&1,6\\
\hline
 &\multicolumn {3}{c}{Axisymmetric Components}  \\
\hline
M$_{\rm D}$/M$_{\rm H}$ $^{\rm 2}$&0.07 &0.09 & 0.1  &  4,8  \\
M$_{\rm B}$/M$_{\rm D}$ & 0.9& 0.4& 0.2 & 5,8 \\
Rotation Velocity $^3$($\kms$)& 320&250 &  170& 7  \\
M$_{\rm D}$ ($10^{10}$M$_\odot$)   &12.8 & 12.14 & 5.10 & 4  \\
M$_{\rm B}$ ($10^{10}$M$_\odot$)  & 11.6& 4.45& 1.02 & $M_{\rm B}/M_{\rm D}$ based \\
M$_{\rm H}$ ($10^{11}$M$_\odot$)     &16.4 &12.5 & 4.85 & $M_{\rm D}/M_{\rm H}$ based \\
Disk Scale-Length ($\kpc$) & 7&5 & 3 & 4,5\\
\hline
&\multicolumn {3}{c}{Constants of the Axisymmetric Components $^{\rm 4}$}  \\
\hline
Bulge (M$_{\rm B}$, b$_1$)$^5$& 5000, 2.5&2094.82, 1.7&440, 1.0&\\
Disk (M$_{\rm D}$, a$_2$, b$_2$)$^5$&5556.03, 7.0, 1.5&5232.75, 5.0, 1.0& 2200, 5.3178, 0.25&\\
Halo (M$_{\rm H}$, a$_3$)$^5$&15000, 18.0&10000, 16.0&2800, 12.0&
\label{tab:parameters}
\enddata

\tablenotetext{1}{The rotation of the spiral arms is clockwise.}
\tablenotetext{2} { Up to 100 kpc halo radius.} 
\tablenotetext{3}{$V_{max}$.}
\tablenotetext{4} { In galactic units, where a galactic mass unit $= 2.32 \times10^7$ M$_\odot$ and a galactic distance unit = kpc.} 
\tablenotetext{5}{b$_1$, a$_2$, b$_2$, and a$_3$ are scale lengths.}

\tablerefs{ (1)~Grosb\o l \& Patsis 1998;
            (2)~Drimmel \et 2000; Grosb\o l \et 2002; Elmegreen \& Elmegreen 2014;
            (3)~Kennicutt 1981; 
            (4)~Pizagno \et 2005;
            (5)~Weinzirl \et 2009;
            (6)~Patsis \et 1991; Grosb\o l \& Dottori 2009; 
                 Egusa \et 2009; Fathi \et 2009; Gerhard 2011;
	    (7)~Brosche 1971; Ma \et 2000; Sofue \& Rubin 2001;
            (8)~Block \et 2002;
            (9)~Pichardo \et 2003;
          (10)~Seigar \& James 1998; Seigar \et 2006.
}

\end{deluxetable} 

In our models, regarding the radial extent of the spiral arms, we
consider as their initial and final galactocentric radii the inner
Lindblad resonance (ILR) and the corotation resonance (CR),
respectively. This is based on theoretical studies of orbital
self-consistency of spiral arms (Contopoulos \& Grosb\o l 1986, 1988;
Patsis \et 1991; Pichardo \et 2003).  In Table \ref{tab:resonances} we
present the positions of the main resonances ILR, 4/1, and CR for
normal spiral galaxies (from early to late types). With a clockwise
rotation for the disk, we assume $\Omega_{\rm p}$ between $10$ and
$60 \kmskpc$, independently of the Hubble type.

\begin{deluxetable} {c| ccc| ccc| ccc}
\tablecolumns{10}

\tabletypesize{\small}
\tablewidth{0pt}
\tablecaption{Resonance Positions}

\tablehead{&\multicolumn{3}{c|}{Sa}&\multicolumn{3}{c|}{Sb}&\multicolumn{3}{c}{Sc}\\
               $\Omega_{\rm p}$ &ILR&4/1&CR&ILR&4/1&CR&ILR&4/1&CR \\
  ($\kmskpc$)& \multicolumn{3}{c|}{$(\kpc)$}&\multicolumn{3}{c|}{$(\kpc)$}&\multicolumn{3}{c}{$(\kpc)$ }
}
\startdata
10 &9.93& 20.25 & 30.23 & 8.62 & 17.74 &26.13 & 4.0&11.32 & 16.78\\
15 &6.46&13.83&20.53 &5.13&12.18&17.88&2.71&7.44&11.49\\
20&4.45 &10.5 &15.66 & 3.52 & 9.22& 13.70 & 2.03&5.35& 8.63 \\
25&3.0&8.44&12.69&2.29&7.34&11.14&1.5&4.11&6.94\\
30 & 3.0 &7.04&10.6 &2.0 &6.04&9.38 &1.5 &3.34& 5.7\\
35&3.0&6.02&9.21&2.0&5.11&8.08&1.5&2.83	&4.8\\
40 & 3.0 &5.24&8.1 & 2.0&4.40&7.0 &1.5 &2.45&4.12\\
50 &3.0 &4.07&6.51 & 2.0&3.42&5.64 &1.5 &1.92& 3.19\\
60 & 3.0&3.21&5.4 & 2.0&2.74& 4.65&1.0& 1.55&2.6

\label{tab:resonances}
\enddata
\end{deluxetable} 

For the mass of the spiral arms, we explore its orbital effect in the
range from 1\% to 10\% of the total disk mass, i.e., $\mu$ between
0.01 and 0.1, independently of the Hubble type.
On the other hand, the spiral arms strength depends mainly
on two parameters, their mass and pitch angle. To measure this
strength and assure that our parameters are within observational
limits, we have employed the $Q_T$ parameter (Combes \& Sanders 1981).
This parameter $Q_T$ is defined as

\begin{eqnarray} \label{q_max}
Q_{\rm T}(R)=F_{T}^{\rm max}(R)/|\langle F_{\rm R} (R) \rangle|, 
\end {eqnarray}

\noindent where F$^{\rm max}_{ T}({\rm R})$ =$|\left(\frac{1}{R}
\ \partial\Phi ({\rm R},\theta)/\partial \theta\right)|_{\rm max}$, is
the maximum amplitude of the tangential force at galactocentric radius
R, and $\langle$F$_{\rm R}$(R)$\rangle$, is the average axisymmetric
radial force at the same radius, derived from the ${\rm m} = 0$ Fourier
component of the gravitational potential.

\begin{figure*}
\includegraphics[trim= 0mm 60mm 2mm 2mm,clip,width=1\textwidth]{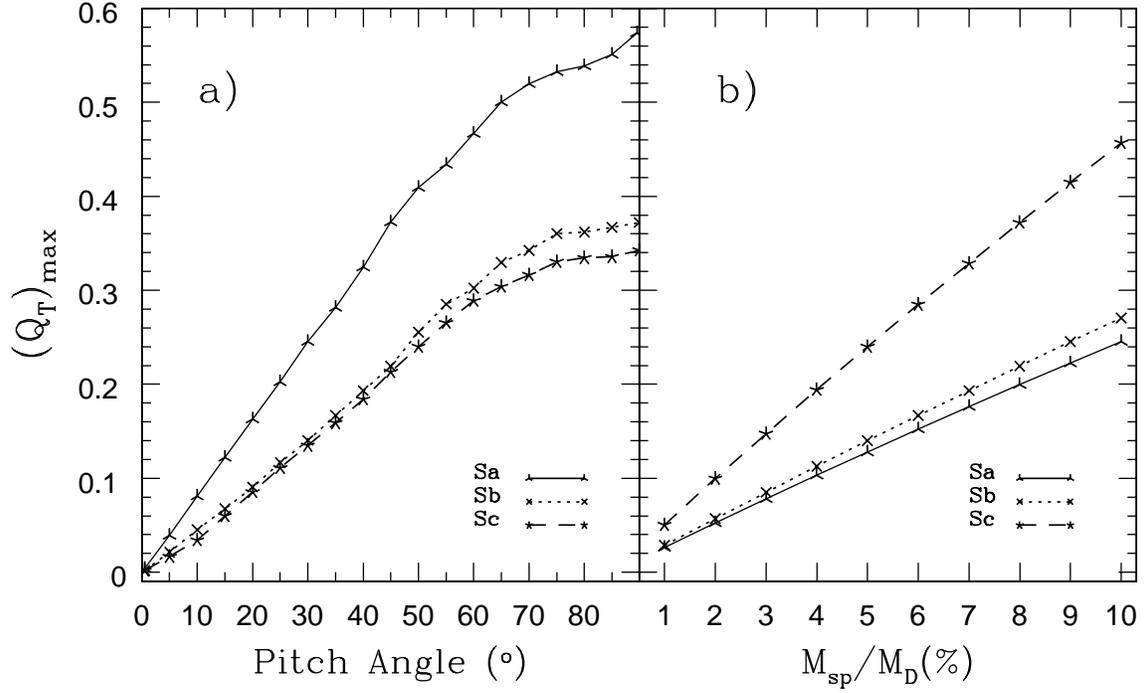}
\caption{Maximum value, (Q$_{\rm T})_{\rm max}$, of the parameter
  Q$_T(R)$. The continuous, dotted, and dashed lines give (Q$_{\rm
    T})_{\rm max}$ for an Sa, Sb, and Sc galaxy, respectively.  In
  $a)$, we show (Q$_{\rm T})_{\rm max}$ vs. pitch angle of the spiral
  arms, $i$, where $\mu$ = M$_{\rm sp}$/M$_{\rm D}$ = 0.1, 0.07, and
  0.05 for Sa, Sb and Sc, respectively. In $b)$, we show (Q$_{\rm
    T})_{\rm max}$ vs. $\mu$, where the pitch angle is $30\deg$,
  $40\deg$ and $50\deg$ for Sa, Sb and Sc, respectively. }
\label{Qm_pa}
\end{figure*}

In Figure \ref{Qm_pa}a we show the maximum value, (Q$_{\rm T})_{\rm
  max}$, of the parameter $Q_T$ for each galaxy type (Sa, Sb and Sc),
as we increase the pitch angle $i$ from $0\deg$ to $90\deg$. For this
figure we employed the maximum plausible value of $\mu$ (i.e. before
chaos dominates the phase space surrounding the families of orbits
that support the pattern; see Section \ref{results}), of 0.1, 0.07,
and 0.05, for Sa, Sb and Sc galaxies, respectively.  In our study,
taking $i$ between $7\deg$ and $30\deg$ (Table \ref{tab:parameters})
and using the maximum limits of the mass of the spiral arms, (Q$_{\rm
  T})_{\rm max}$ is not larger than $\sim 0.25$. From the literature,
reasonable maximum values for $Q_T$ are $\sim0.2$ for early spirals
and $\sim0.3$ for late spirals (Buta \et 2005), therefore, our models
are within observational limits.

In Figure \ref{Qm_pa}b we show (Q$_{\rm T})_{\rm max}$ for each galaxy
type as we increase $\mu$ from 0.01 to 0.1. In this figure we have
considered the value of $i$ before chaos dominates the available phase
space, which corresponds to $\sim30\deg$, $\sim40\deg$ and
$\sim50\deg$, for Sa, Sb and Sc galaxies, respectively (see Section
\ref{results}). For Sa and Sb models, (Q$_{\rm T})_{\rm max}$
increases much slower than for the Sc ones. In early galaxy type
models very massive spiral arms are allowed within this observational
restriction without exceeding observational limits, even for spiral
arms with $\mu$ = 0.1; see for example (Weiner \& Sellwood 1999). For
the Sc type, (Q$_{\rm T})_{\rm max}$ increases much faster with mass,
and for approximately $\mu$ $>$ 0.06 the spiral arms strength
increases ((Q$_{\rm T})_{\rm max}> 0.3$) more than what observations
indicate for this type of galaxies.

\subsection{Tools for the Orbital Analysis}\label{analysis}
    
For the orbital dynamics analysis we employed periodic orbits and
Poincar\'e diagrams. We have also calculated the density response as
in Paper I, using the method of Contopoulos \& Grosb\o l (1986), which
quantifies the support of spiral arms with periodic orbits. This
method to estimate the density response has been widely used in
literature (Contopoulos \& Grosb\o l 1988; Amaral \& Lepine 1997; Yano
\et 2003; Pichardo \et 2003; Voglis \et 2006; Tsoutsis \et 2008;
P\'erez-Villegas \et 2012, 2013; Junqueira \et 2013). We computed
between 40 and 60 periodic orbits for each galactic model. The density
response is defined as the regions where periodic orbits crowd
producing a density enhancement. The position of the maximum density
response along each periodic orbit is calculated, and with these
positions the locus formed by the maxima of density response is found
and compared with the position of the imposed spiral locus
(i.e. PERLAS). The method implicitly considers a small and
  variable dispersion since it studies a region where the flux is
  conserved. Additionally, we estimated the average density response
  around each one of these maxima response. In order to do that, we
  took a circular vicinity, and compared the density response with the
  imposed density (this is the sum of the axisymmetric disk density on
  the galactic plane and the central density of the spiral arms).
 
Regarding the Poincar\'e diagrams, these are constructed in the plane
$(x',v_x')$, in the non-inertial reference system that rotates with
the spiral arms. The $x'$ axis points toward the direction of the line
where the spiral arms begin in the inner galactic region. Poincar\'e
diagrams have two regions: the prograde region, where the stars move
in the same direction of rotation of the spiral arms, and the
retrograde region, where the stars move in opposite direction to the
spiral arms rotation. These regions (prograde and retrograde) were
defined in the non-inertial frame where the spiral arms are at
rest. In our models the rotation of the spiral arms is clockwise,
thus, the left side of the diagram is prograde (launching orbits with
$x'<0, v'_y>0$), and the right side is retrograde (with $x'>0,
v'_y>0$). Ordered orbits are shown as one-dimensional curves, periodic
orbits as a few set of dots (one single dot represents the strongest
periodic orbits, surrounded by other ordered, quasi-periodic and
periodic orbits of lower order), and chaotic orbits are seen as
scattered sets of points. The chaotic regions may surround periodic
orbits, and confined chaotic orbits is able to support large-scale
structures such as spiral arms (Patsis \& Kalapotharakos 2011;
Kaufmann \& Contopoulos 1996; Contopoulos \& Grosb\o l 1986). However,
an excess of chaos would start making it difficult for orbits to
support structures, until chaos is not confined anymore, destroying
large scale patterns (chaos in this case penetrates up to the $X_1$
periodic orbits regions). Large-scale structures are not expected to
arise from systems fully dominated by chaos (Voglis et al. 2006). The
most interesting part of the phase space diagrams is the prograde
region where the great majority of stars are moving in spiral
galaxies. Chaos is generated in this region, mostly due to resonance
overlapping (Martinet 1974; Athanassoula \et 1983 and references
therein). In the retrograde region, the resonances are very separated,
thus the production of chaos is almost null. Each Poincar\'e diagram
contains 50 orbits, distributed between the prograde and retrograde
regions, with 300 points each (points correspond to the numbers of
periods).  For more details about this methodology (periodic orbits,
density response and phase-space diagrams), see Paper I.

\section{Orbital Study of Ordered and Chaotic Behavior} \label{results}

With the methodology described in the previous section, we have
carried out a detailed orbital study with periodic orbits, density
response calculations, and phase space diagrams. With all these tools
we try to determine whether limiting values to different structural
and dynamical parameters of normal spiral galaxies can be
established. In Paper I we found two limits to the pitch angle $i$ for
normal spiral galaxies (Sa, Sb and Sc); in the present study we set
limits to combinations of the spiral arms-to-disk mass ratio, $\mu$,
the angular velocity of the spiral arms, $\Omega_{\rm p}$, combined
with the pitch angle, $i$, in order to seek for plausible long-term
spiral galactic models. By plausible and long-lasting in this context,
we mean spiral arms fully supported by periodic orbits and moderate
production of chaotic behavior.
 
To measure observationally both the mass and angular velocity of
spiral arms is not an easy task. We have tried instead to constrain
these parameters through the orbital support of the spiral arms with
periodic orbits (ordered behavior), and with the study of chaotic
behavior with Poincar\'e diagrams, by searching for a limit before
chaos dominates the available phase space and destroys periodic
orbits.

We present in this section a family of orbitally plausible
long-lasting spiral galactic potentials, and provide optimal ranges
for the parameters $i$, $\Omega_{\rm p}$, and $\mu$.

\subsection{Analyzing the Effect of the Spiral Arms Mass: Ordered and Chaotic Behavior} 
\label{mass}

We performed an exhaustive study of periodic orbits for different
morphological galactic types.  With the maps of periodic orbits we
found the position of the maximum density response along each periodic
orbit, and in order to analyze some orbital self-consistency of the
spiral arms, these positions were compared with the center of the
imposed spiral pattern. If the imposed spiral arms are supported by
the maxima of density response, then these arms are stable and are of
a long-lasting nature. If this condition is not satisfied, the spiral
arms might be rather explained as transient structures.

For each morphological type, we used the corresponding axisymmetric
background potential, based on the parameters presented in Table
\ref{tab:parameters}. In order to dilucidate the relative importance
of the different parameters, we present in this section several
examples first. $\mu$ has been varied from 0.01 to 0.1, and the
representative employed pitch angles $i$ are taken as $10\deg$,
$15\deg$ and $20\deg$, for Sa, Sb and Sc galaxies,
respectively. Additionally, in our computations we have varied
slightly $\Omega_{\rm p}$ in each galactic type.

\begin{figure*}

\includegraphics[width=.95\textwidth]{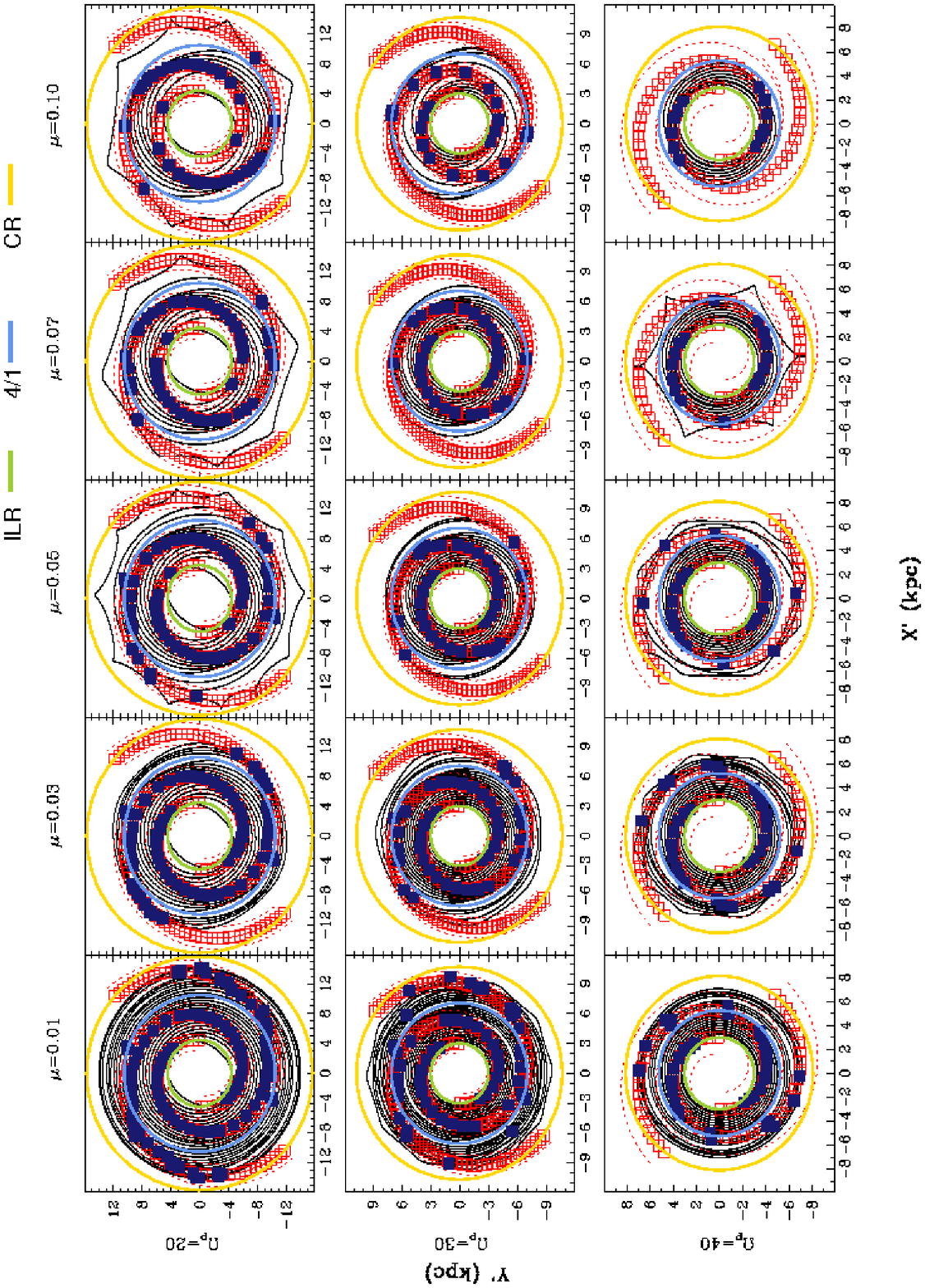}
\caption{\small Periodic orbits (black curves), density response
  maxima (filled squares), and the imposed spiral arms locus (open
  squares and dotted lines mark the width of spiral arms) for the
  three-dimensional spiral arms model of an Sa galaxy with a pitch
  angle $i$ = $10\deg$. The values of $\mu$ = $M_{\rm sp}/M_{\rm D}$
  and the angular speed of the spiral arms, $\Omega_{\rm p}$ in units
  of $\kmskpc$, are given at the top and left, respectively.}
\label{p_mass_sa_pa10}
\end{figure*}

In Figure \ref{p_mass_sa_pa10}, for an Sa galaxy with $i$ = $10\deg$,
we show in the galactic plane $x',y'$ the maxima of density response,
with filled squares, which correspond to crowding regions of periodic
orbits (black curves) that produce density enhancements; the center of
the imposed spiral arms potential (PERLAS model) is shown with open
squares and the dotted lines mark the width of spiral arms. From top
to bottom panels (the $x'$ axis is at the bottom), $\Omega_{\rm p}$
lies in the interval $[20,40]$ $\kmskpc$, and the values of $\mu$ are
marked at the top. In each panel we show with red, blue, and yellow
circles the positions of the resonances ILR, 4/1 and CR; see Table
\ref{tab:resonances}. This figure shows that the maxima of density
response coincide well (within $3\deg$ difference) with the imposed
spiral arms, but the extent of this density support reduces
significantly for the largest values of $\mu$, reaching only up to the
4/1 resonance if $\mu$ is around 0.1. Also, the density support
diminishes strongly if $\Omega_{\rm p}$ increases.

\begin{figure*}
\includegraphics[width=.95\textwidth]{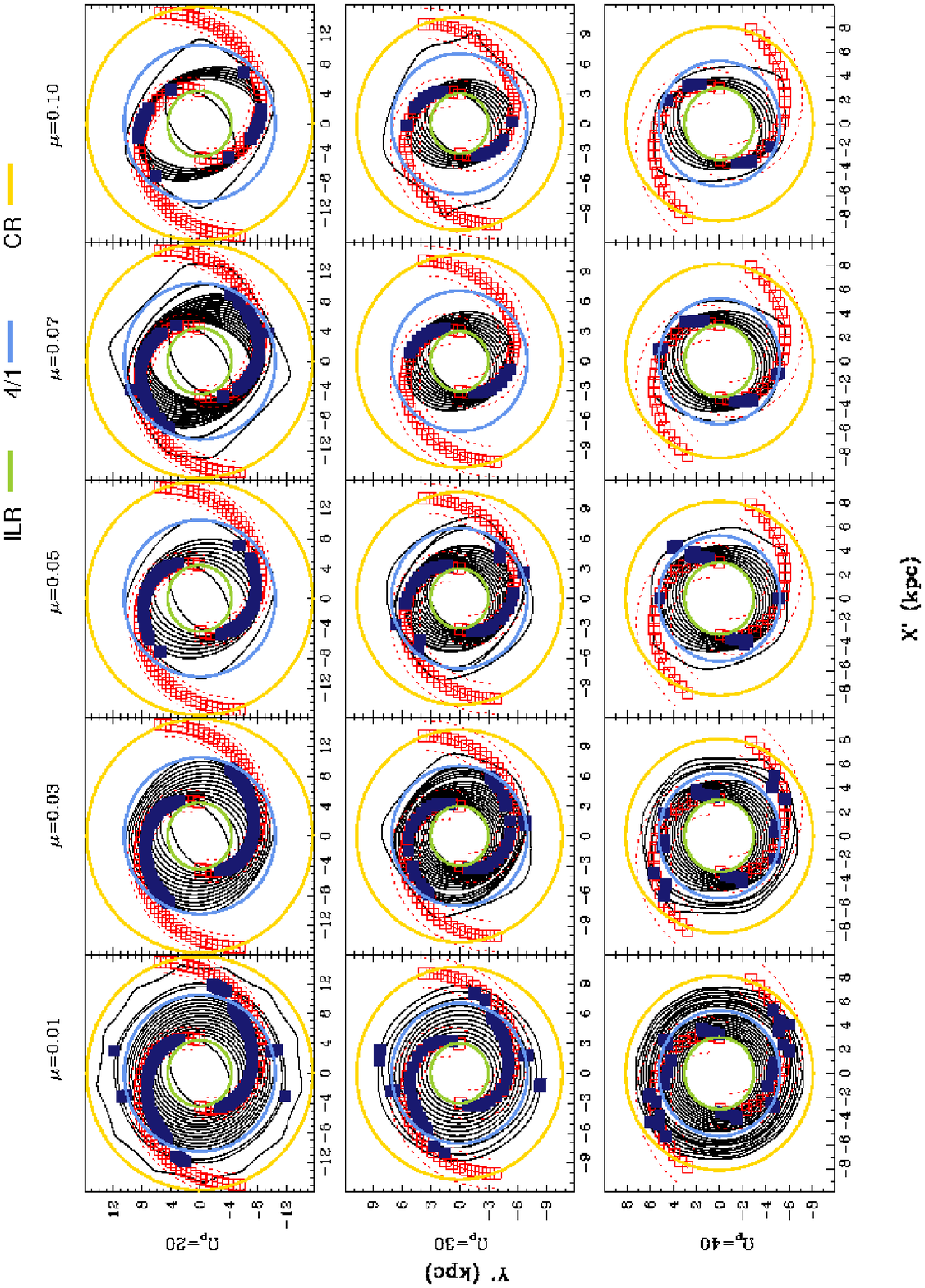}
\caption{As in Figure \ref{p_mass_sa_pa10}, here with $i$ = $20\deg$.} 
\label{p_mass_sa_pa20}
\end{figure*}

In Figure \ref{p_mass_sa_pa20} we consider $i$ = $20\deg$ in an Sa
type galaxy. In this case the orbital dynamics is strongly affected.
The density response systematically forms spiral arms with a smaller
pitch angle than the imposed $20\deg$, and with a reduced radial
extent compared with the case $i$ = $10\deg$. The partial density
support is destroyed when $\Omega_{\rm p}$ increases. Spiral arms with
this strong forcing in a galaxy, might be better explained as
transient structures. In Paper I we found that the regime where the
spiral arms of Sa galaxies are transient occurs when $i$ $\gtrsim$
$15\deg$.

\begin{figure*}
\includegraphics[width=.95\textwidth]{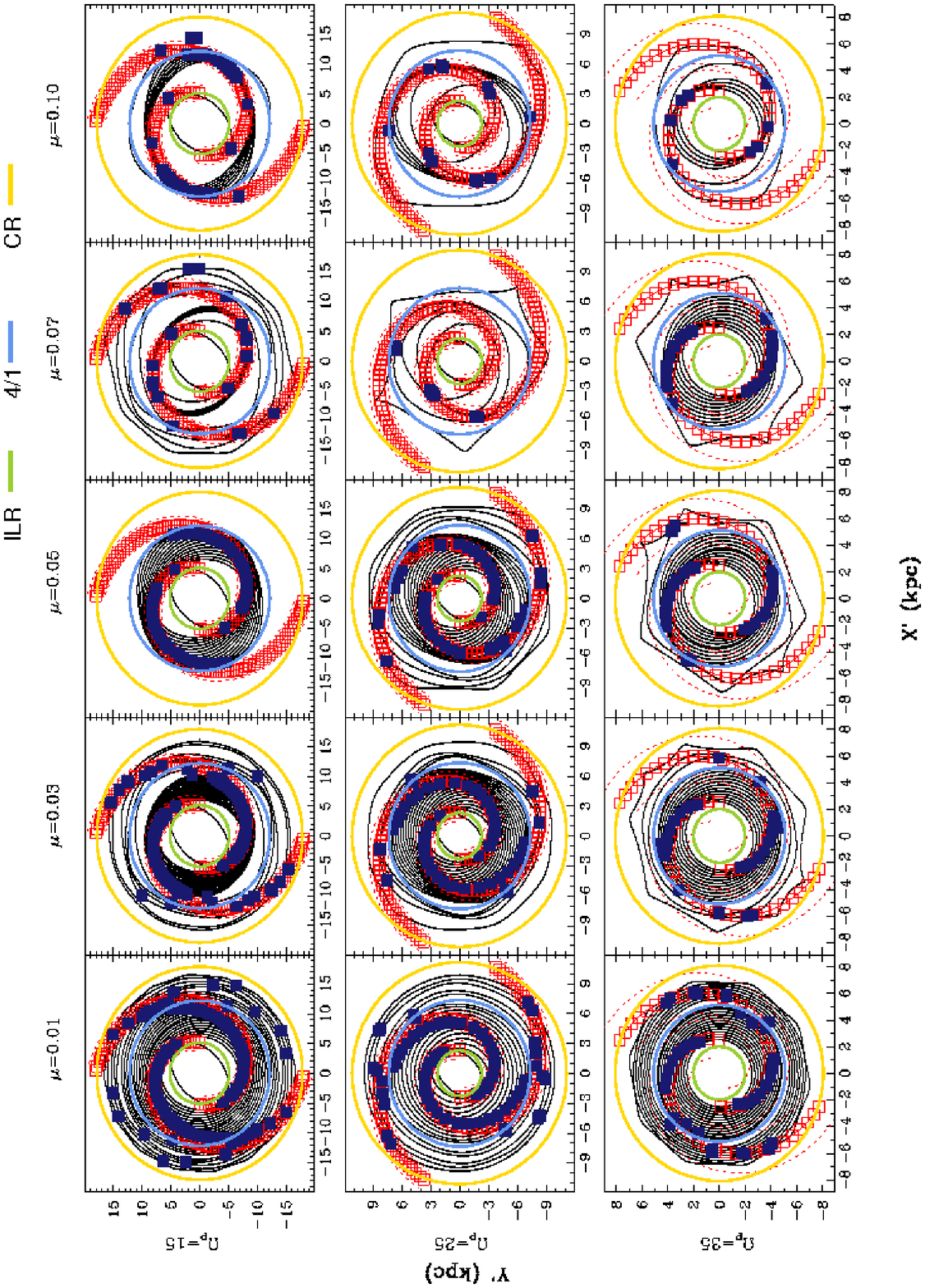}
\caption{\small Periodic orbits (black curves), density response
  maxima (filled squares), and the imposed spiral arms locus (open
  squares and dotted lines mark the width of spiral arms) for the
  three-dimensional spiral arms model of an Sb galaxy with a pitch
  angle $i$ = $15\deg$. The values of $\mu$ and $\Omega_{\rm p}$ are
  given at the top and left, respectively.}
\label{p_mass_sb_pa15}
\end{figure*}

\begin{figure*}
\includegraphics[width=.95\textwidth]{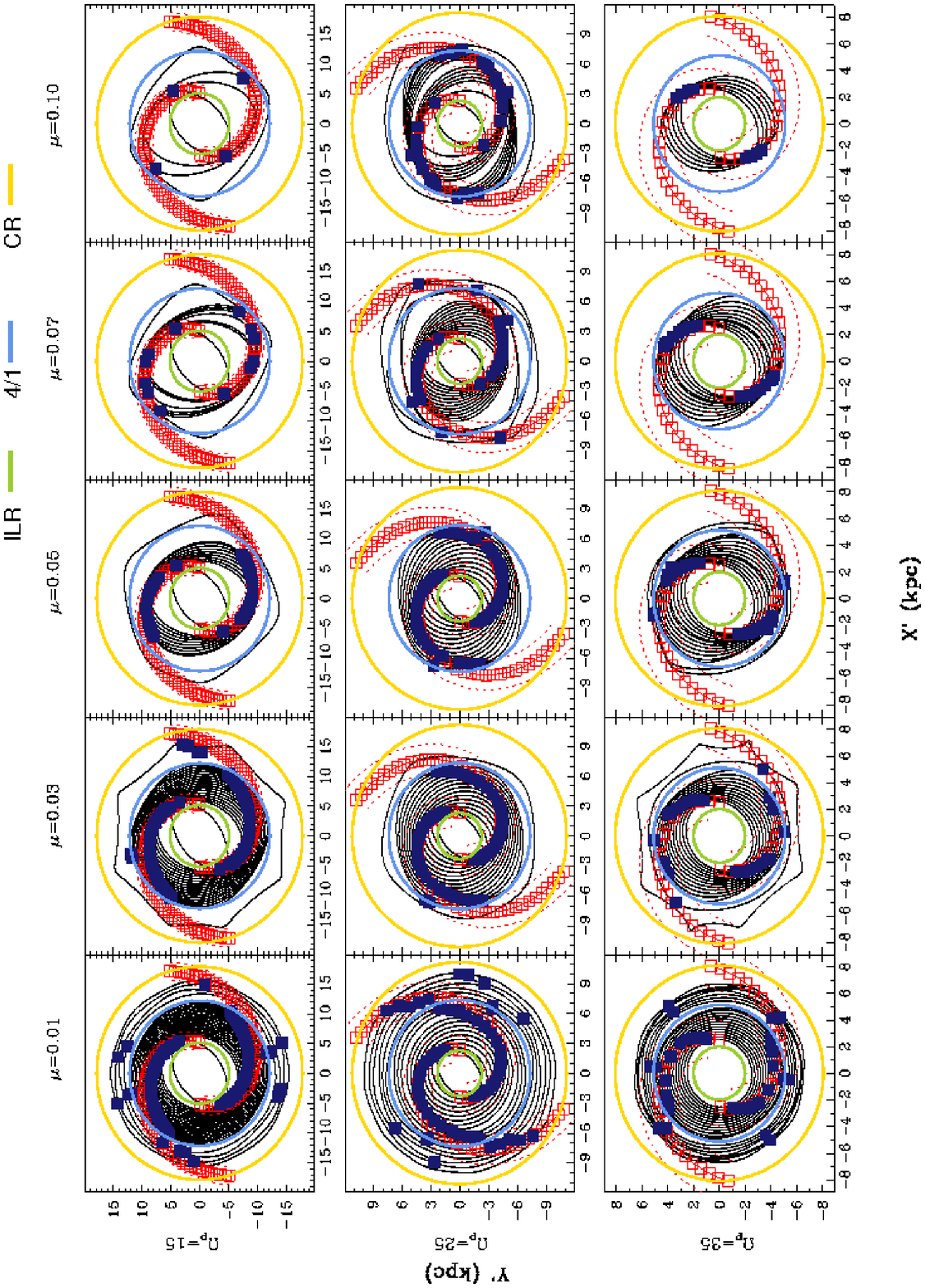}
\caption{As in Figure \ref{p_mass_sb_pa15}, here with $i$ = $20\deg$.} 
\label{p_mass_sb_pa20}
\end{figure*}

Figures \ref{p_mass_sb_pa15} and \ref{p_mass_sb_pa20} show our results
for an Sb galaxy with pitch angles $i$ = $15\deg$ and $i$ = $20\deg$,
respectively. In these cases $\Omega_{\rm p}$ lies in the interval
$[15,35]$ $\kmskpc$. These figures show a similar behavior to the case
of an Sa galaxy, but it was harder to obtain a reasonable density
support for larger values of $\mu$. With the greater value $i$ =
$20\deg$, the resulting response pitch angle is slightly smaller than
the imposed $20\deg$.

\begin{figure*}
\includegraphics[width=.95\textwidth]{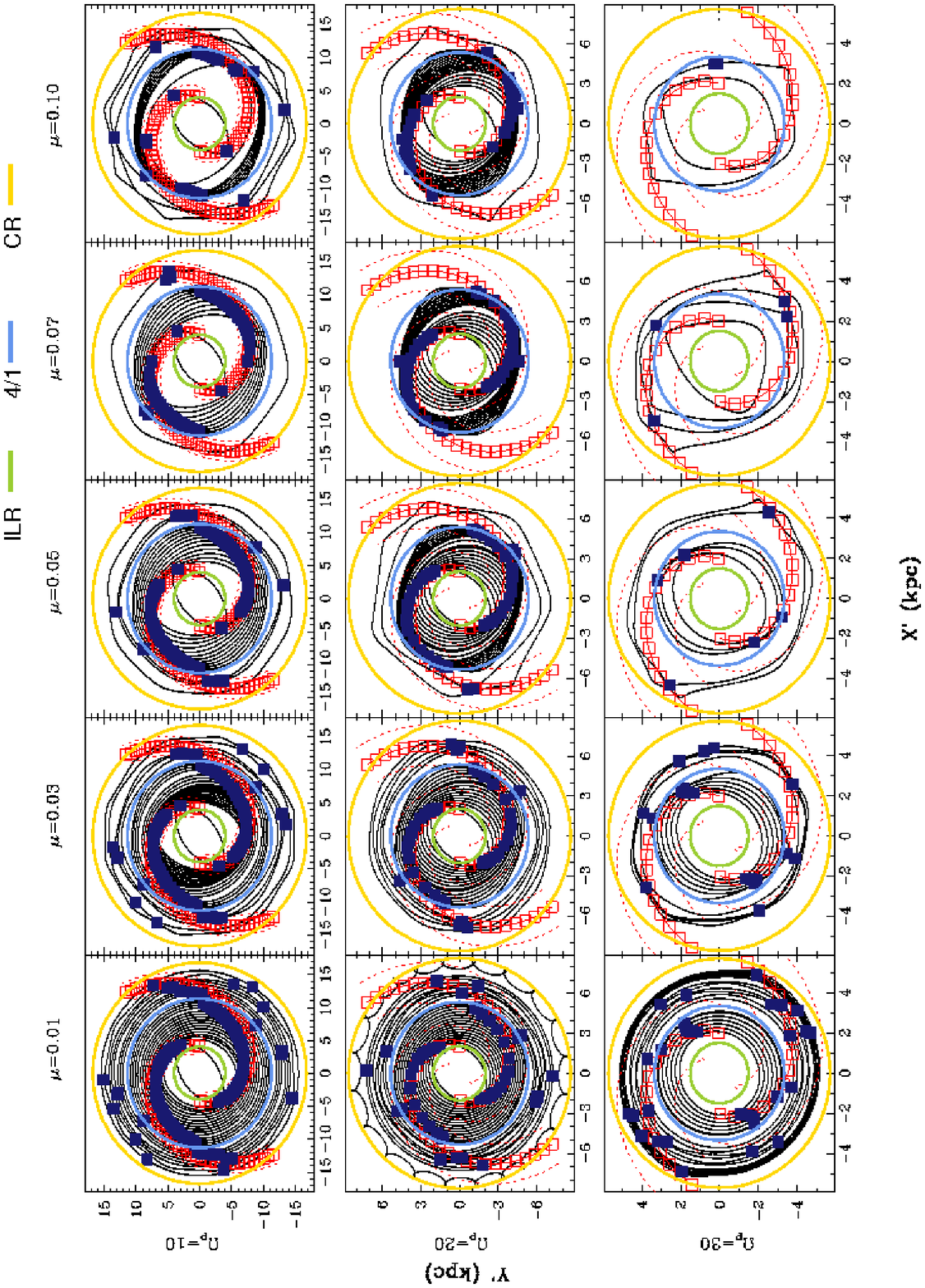}
\caption{\small Periodic orbits (black curves), density response
  maxima (filled squares), and the imposed spiral arms locus (open
  squares and dotted lines mark the width of spiral arms) for the
  three-dimensional spiral arms model of an Sc galaxy with a pitch
  angle $i$ = $20\deg$. The values of $\mu$ and $\Omega_{\rm p}$ are
  given at the top and left, respectively.}
\label{p_mass_sc_pa20}
\end{figure*}

\begin{figure*}
\includegraphics[width=.95\textwidth]{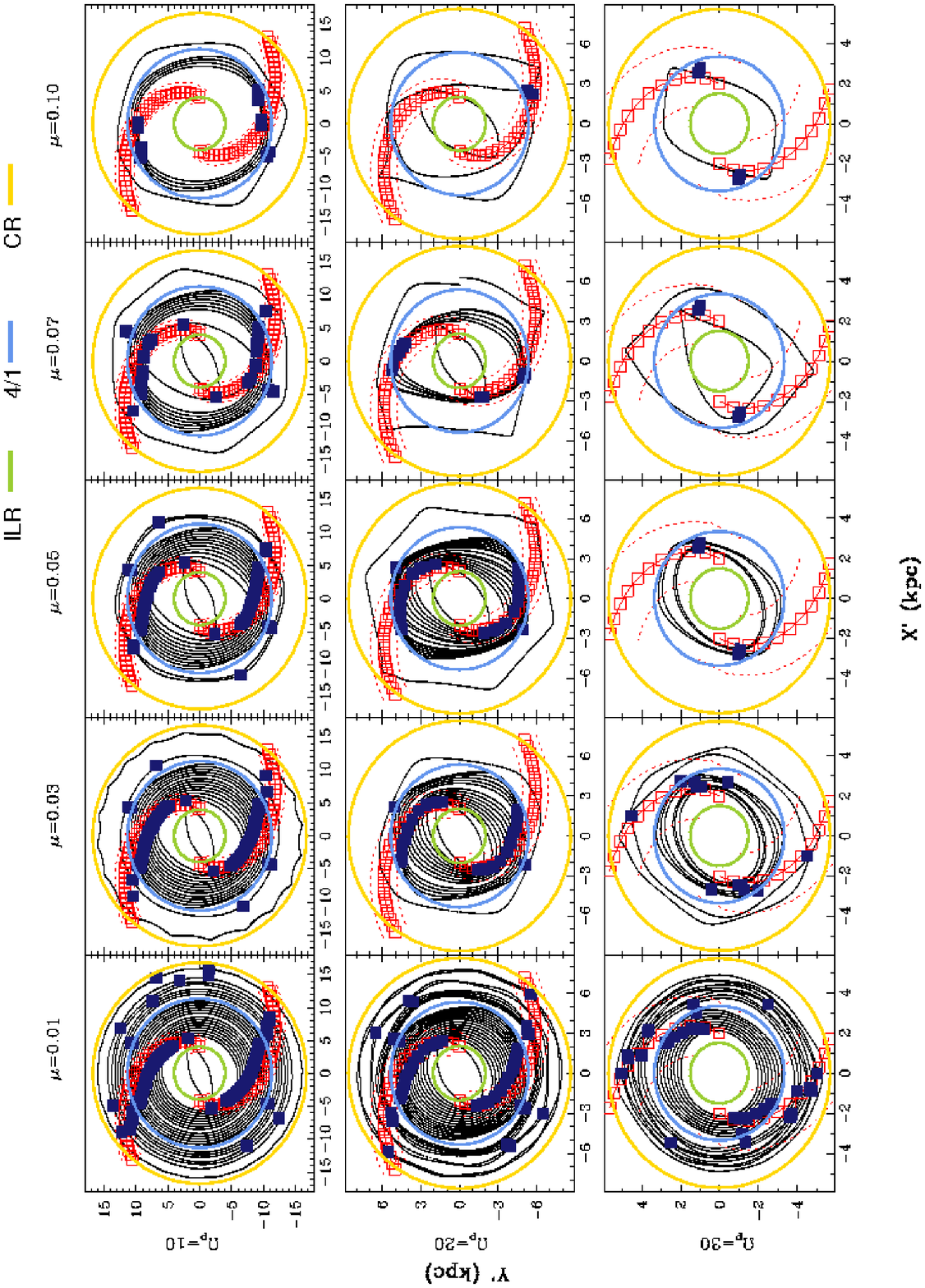}
\caption{As in Figure \ref{p_mass_sc_pa20}, here with $i$ = $30\deg$.}
\label{p_mass_sc_pa30}
\end{figure*}

Our results for an Sc galaxy are shown in Figures \ref{p_mass_sc_pa20}
and \ref{p_mass_sc_pa30}, with pitch angles $i$ = $20\deg$ and $i$ =
$30\deg$, respectively. $\Omega_{\rm p}$ lies in the interval
$[10,30]$ $\kmskpc$. We do not find a density support toward the
greatest values of $\mu$ and $\Omega_{\rm p}$, and for their smallest
values the response density shows a pitch angle smaller than the
imposed one. The radial extent of this response shortens compared with
that obtained in Sa and Sb types.
 
\begin{figure*}
\includegraphics[width=.95\textwidth]{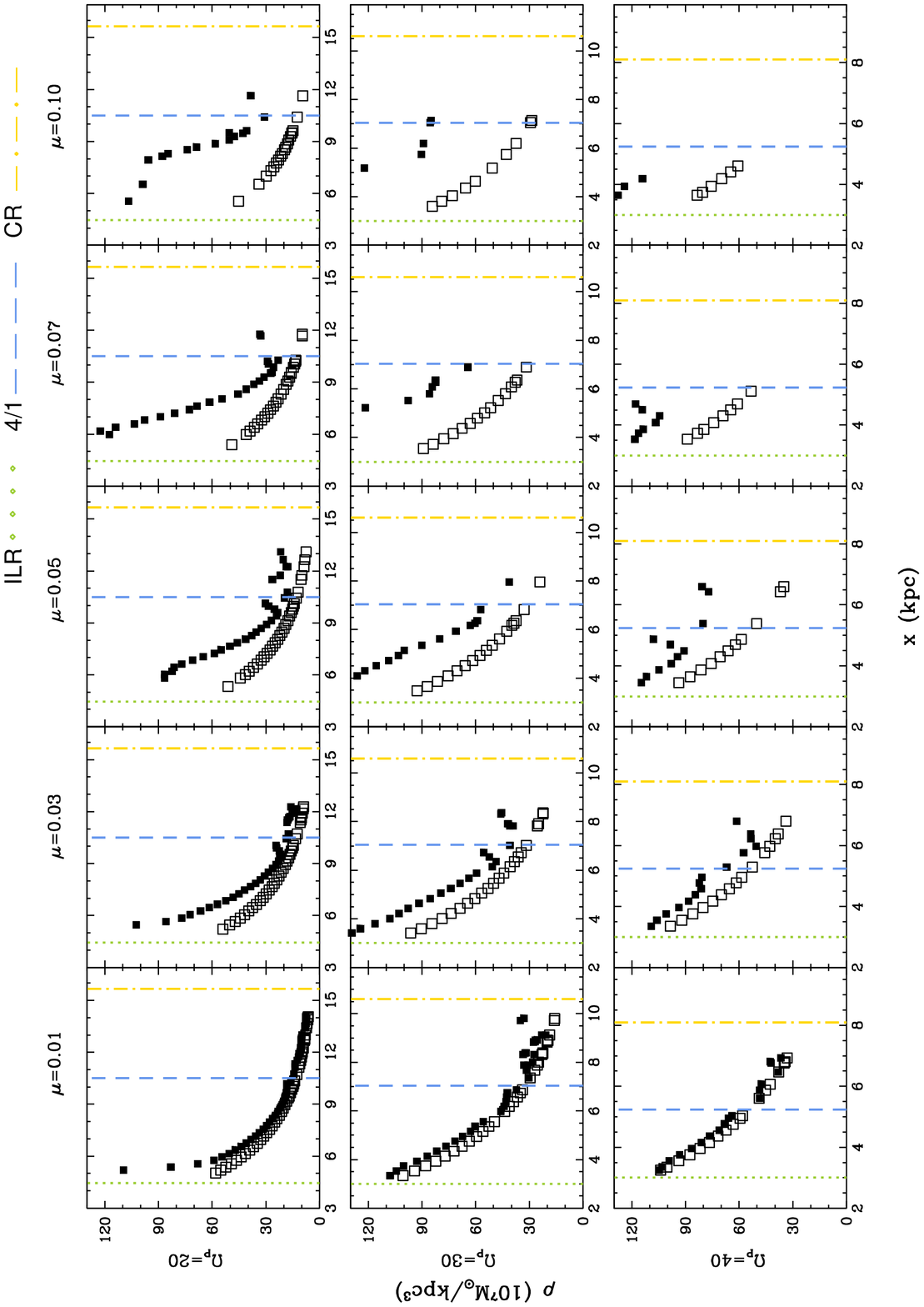}
\caption{ \small Filled squares are the density response of spiral
  arms for an Sa galaxy, and open squares represent the imposed
  density with a pitch angle $i$ = $10\deg$. The values of $\mu$ and
  $\Omega_{\rm p}$ are given at the top and left, respectively. The
  dotted, dashed and dot-dashed lines show the ILR position, 4/1
  resonance position and CR position, respectively.}
\label{r_mass_sa_pa10}
\end{figure*}

\begin{figure*}
\includegraphics[width=.95\textwidth]{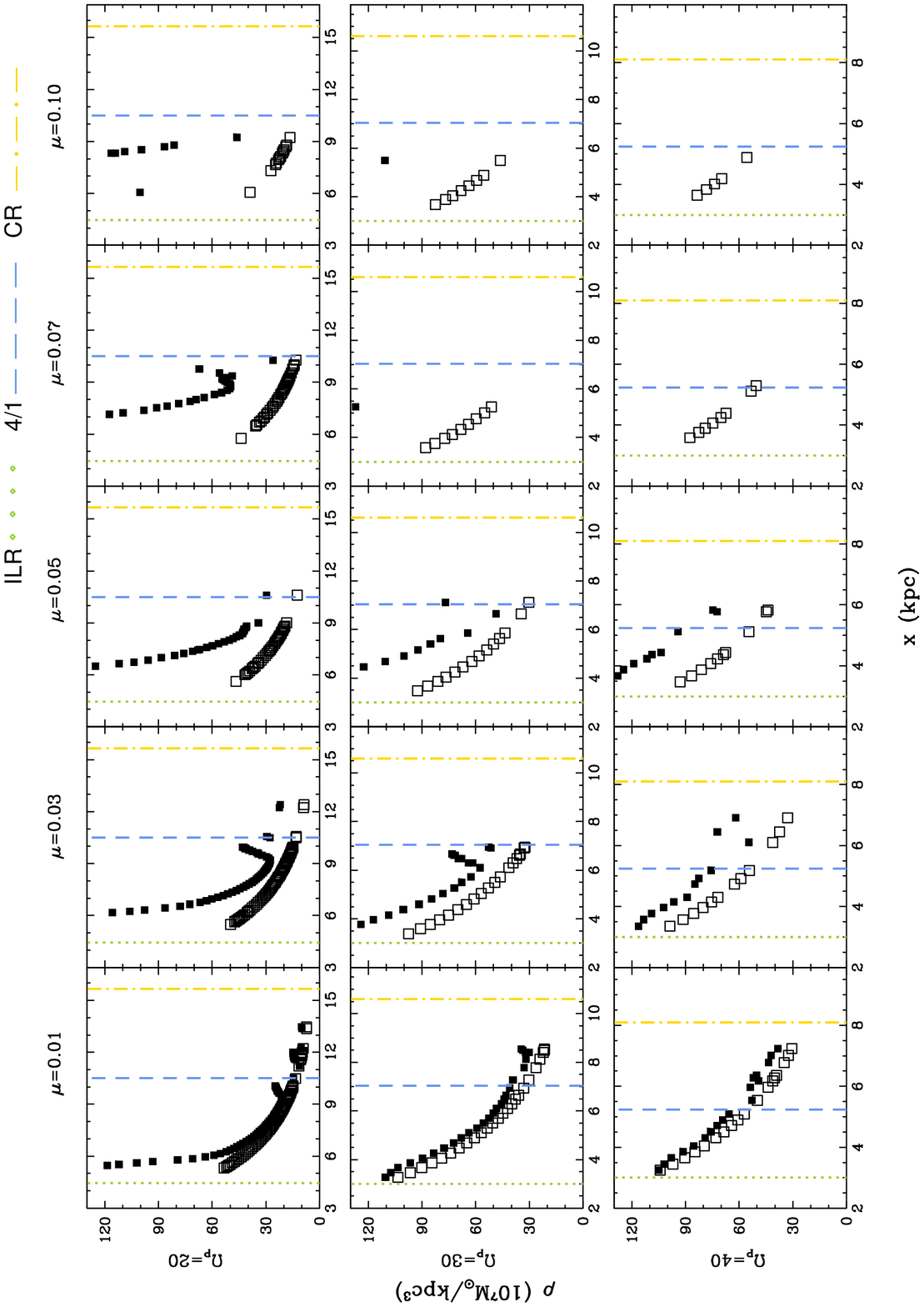}
\caption{ As in Figure \ref{r_mass_sa_pa10}, here with $i$ = $20\deg$.}
\label{r_mass_sa_pa20}
\end{figure*}

\begin{figure*}
\includegraphics[width=0.95\textwidth]{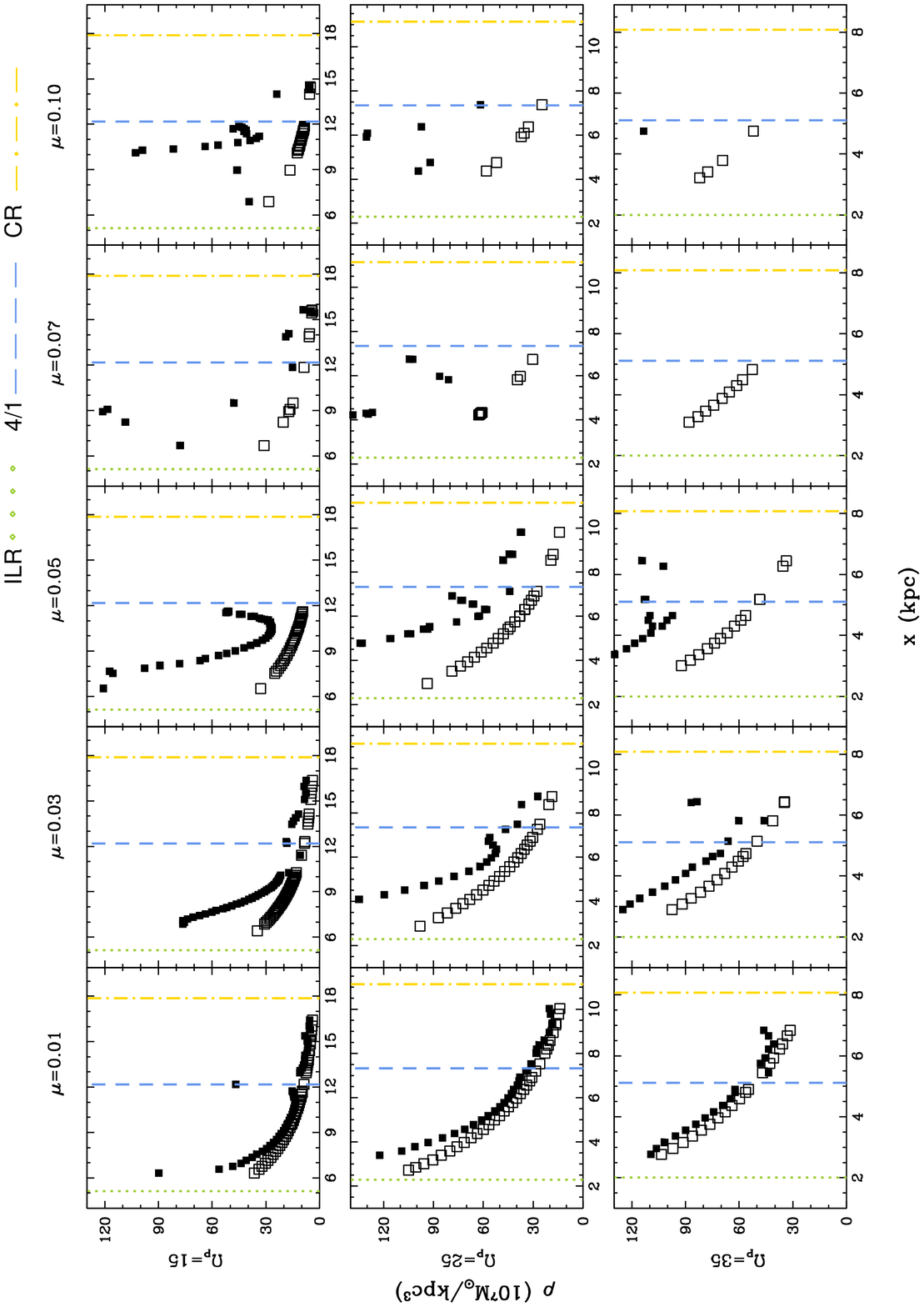}
\caption{ \small Density response diagrams. Filled squares are the
  density response of spiral arms for an Sb galaxy, and open squares
  represent the imposed density with a pitch angle $i$ = $15\deg$. The
  values of $\mu$ and $\Omega_{\rm p}$ are given at the top and left,
  respectively. The dotted, dashed and dot-dashed lines show the ILR
  position, 4/1 resonance position and CR position, respectively.}
\label{r_mass_sb_pa15}
\end{figure*}

\begin{figure*}
\includegraphics[width=0.95\textwidth]{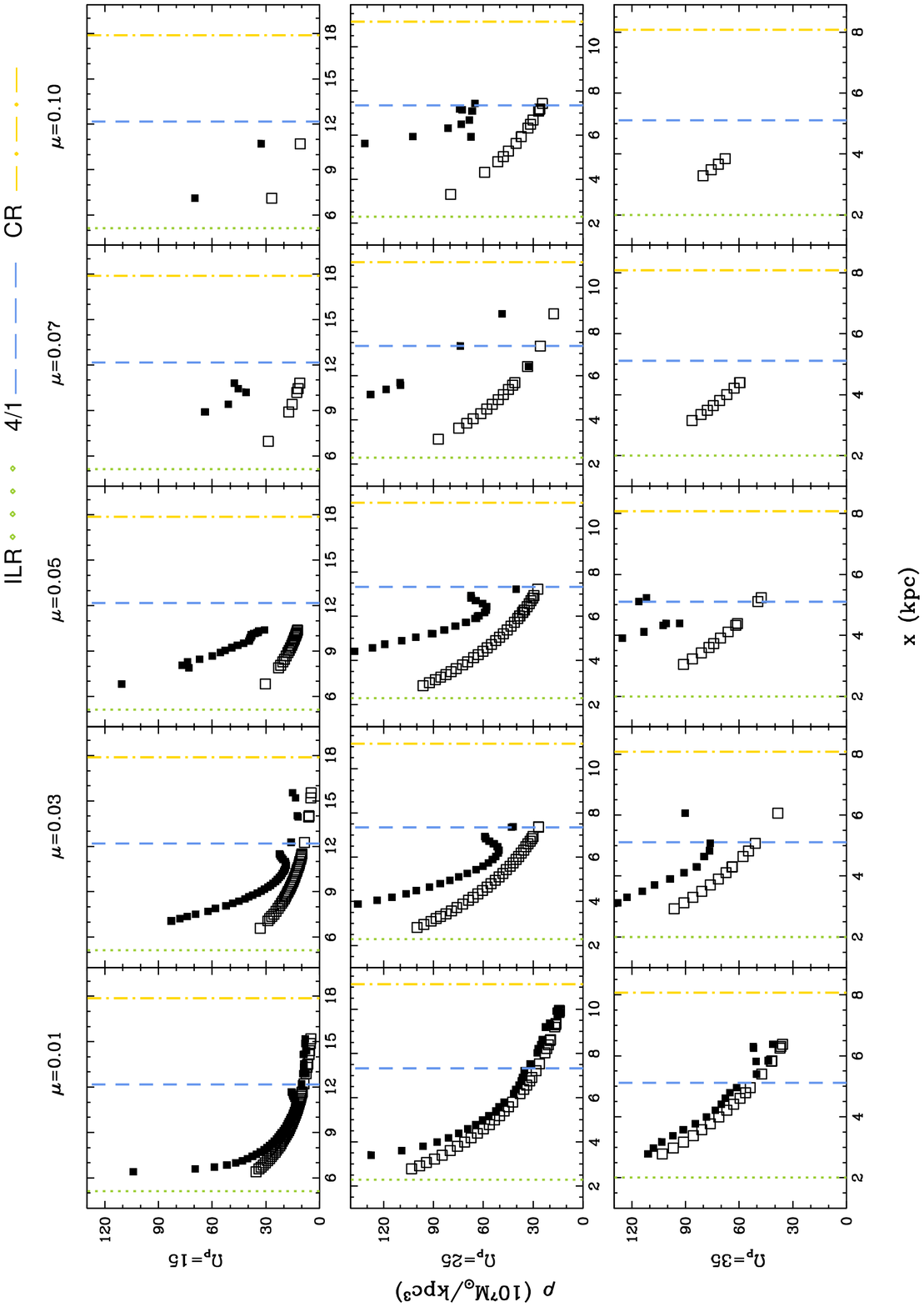}
\caption{ As in Figure \ref{r_mass_sb_pa15}, here with $i$ = $20\deg$.}
\label{r_mass_sb_pa20}
\end{figure*}

\begin{figure*}
\includegraphics[width=0.95\textwidth]{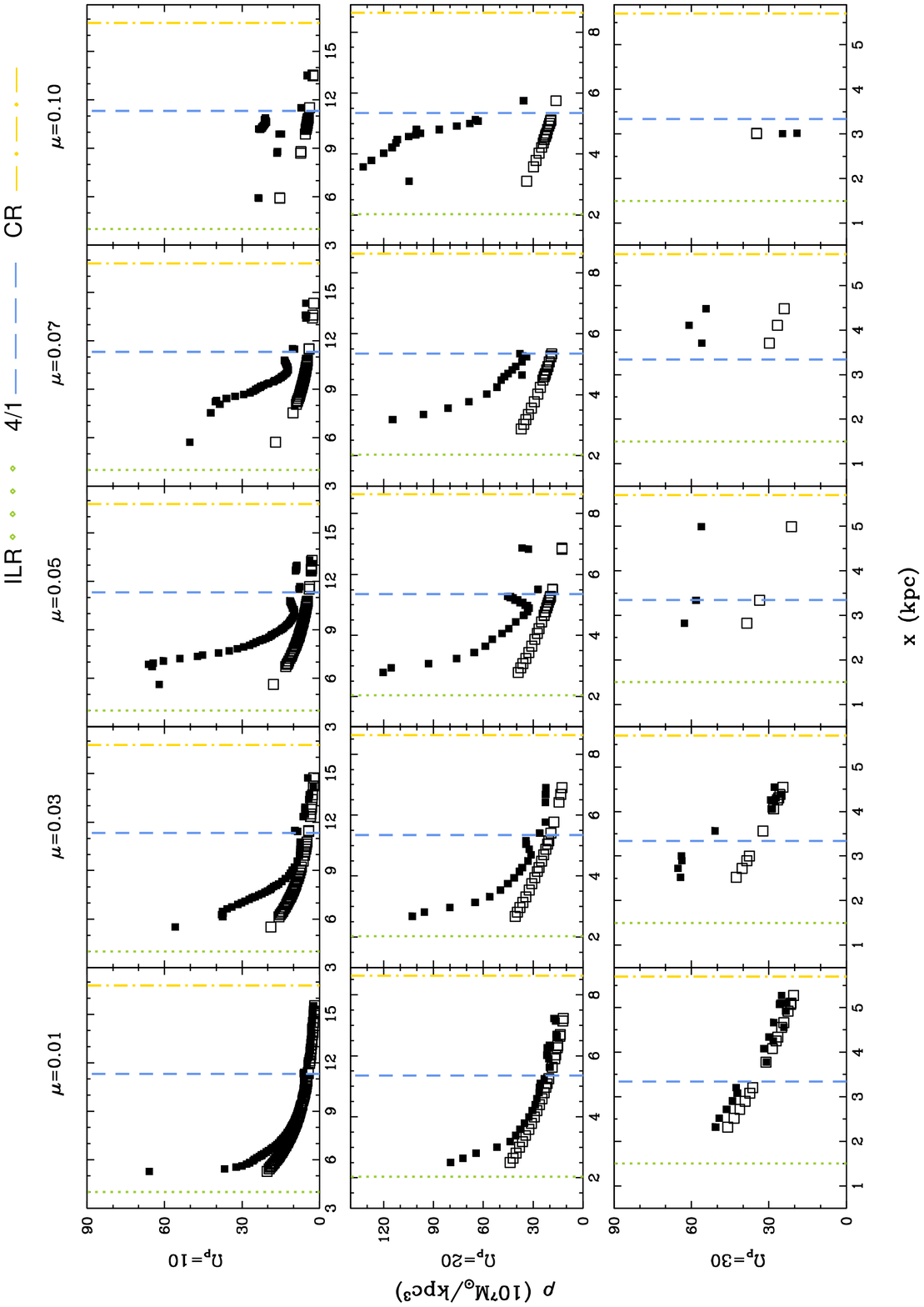}
\caption{ \small Density response diagrams. Filled squares are the
  density response of spiral arms for an Sc galaxy, and open squares
  represent the imposed density with a pitch angle $i$ = $20\deg$. The
  values of $\mu$ and $\Omega_{\rm p}$ are given at the top and left,
  respectively. The dotted, dashed and dot-dashed lines show the ILR
  position, 4/1 resonance position and CR position, respectively.}
\label{r_mass_sc_pa20}
\end{figure*}

\begin{figure*}
\includegraphics[width=0.95\textwidth]{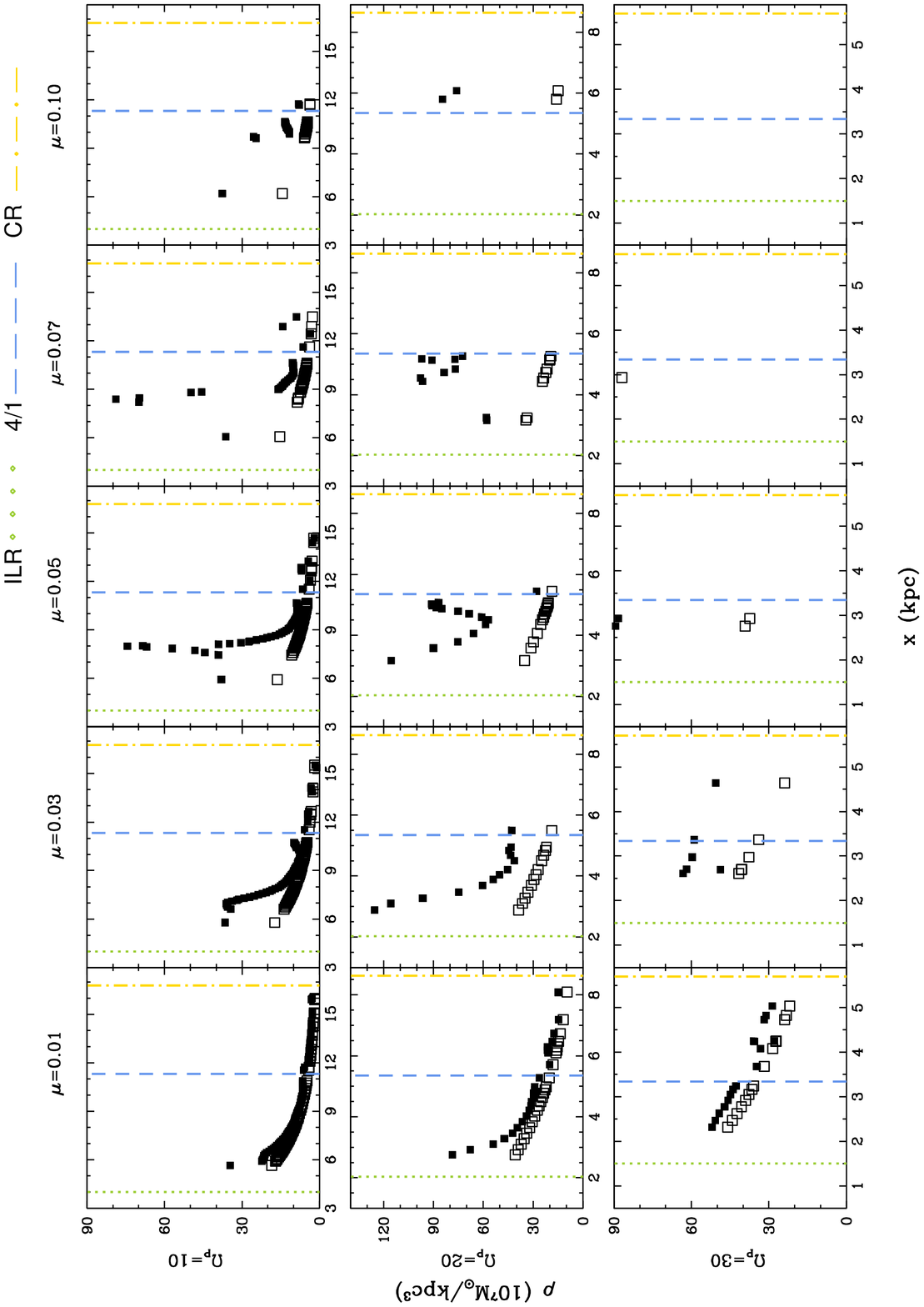}
\caption{ As in Figure \ref{r_mass_sc_pa20}, here with $i$ = $30\deg$.}
\label{r_mass_sc_pa30}
\end{figure*}

In order to complement and to reinforce the results obtained by
  the construction of periodic orbits, in Figures \ref{r_mass_sa_pa10}
  -\ref{r_mass_sc_pa30}, we have compared the spiral arm density
  response (filled squares) with the spiral arms imposed density
  (PERLAS, open squares). Each mosaic of density response (Figures
  \ref{r_mass_sa_pa10} -\ref{r_mass_sc_pa30}) corresponding to each
  periodic orbit mosaic (Figures \ref{p_mass_sa_pa10}
  -\ref{p_mass_sc_pa30}). In Figure \ref{r_mass_sa_pa10}, we present
  densities for an Sa galaxy with $i$ = $10\deg$. As the maximum
  density response was shown in Figure \ref{p_mass_sa_pa10}, this
  figure presents that for $\mu$ up to $ \sim 0.05$, the density
  response fits well with the imposed density. Figure
  \ref{r_mass_sa_pa20} shows also a Sa galaxy, but with $i$ =
  $20\deg$, in this case the spiral arms are stronger, and we see that
  density response fits to imposed density with a smaller $\mu$.
  Therefore, if the pitch angle increases, the allowed mass in spiral
  arms should be smaller, in order to maintain the orbital support and
  the density response fits better to the imposed density.

 Figures \ref{r_mass_sb_pa15} and \ref{r_mass_sb_pa20} show
  density response for an Sb galaxy with pitch angles $i$ = $15\deg$
  and $i$ = $20\deg$, respectively. In these figures we see a similar
  behavior than for Sa galaxies. The density response in this case,
  fits the imposed density up to $\mu \sim 0.03$, but if the pitch
  angle increases, the value of $\mu$ is affected.
 
 Figures \ref{r_mass_sc_pa20} and \ref{r_mass_sc_pa30} show
  densities for an Sc galaxy with pitch angles $i$ = $20\deg$ and $i$
  = $30\deg$, respectively. In these figures we see a similar behavior
  than in Sa and Sb galaxies. We can notice that the density response
  fits to imposed density up to $\mu \sim 0.03$. For larger values
  of $\mu$ and $\Omega_{\rm p}$, the density support is not found.

Figures \ref{r_mass_sa_pa10} -\ref{r_mass_sc_pa30}, show that the response is compatible with the imposed densities up to a certain limit in mass. The larger the force of the spiral arms, the stronger the response relative to the imposed one. It is worth noticing that, this over-response does not indicate that the model is inconsistent as long as the response is in phase. Rather, it would indicate a growing mode, which would be probably damped by an increase in velocity dispersion, if feedback were included in a totally self-consistent model.

Now, in order to study the chaotic behavior, we produced a comprehensive
study of Jacobi energy ($E_J$) families in phase space, from very
  bounded orbits (the inner part of galaxy) to the outer galaxy, even
  in some cases passing the corotation barrier. In the Poincar\'e
diagrams we varied the spiral arms mass, and their angular velocity
and pitch angle, as we did with the periodic orbital study. The
analysis of the chaotic behavior is relevant because it can provide
constraints to the maximum values of some important parameters of
galaxies (pitch angle or spiral arms masses, for example). With this
study, we find a limit to the spiral arms mass, for which chaos
becomes pervasive dominating the available phase space and destroying
all periodic orbits as well as the ordered orbits surrounding them.

We present a set of Poincar\'e diagrams for each morphological type.
In our experiments, assorted spiral arms masses, pitch angles and
angular velocities are tested. We explored a comprehensive set of
  $E_J$ families, from energies representing the most bounded orbits
  (galactic centers) to the corotation barrier and beyond to cover the
  total extension of the spiral arms. The values presented in the
  mosaics of Poincar\'e diagrams correspond approximately to the CR
  position in each case (that represent the most extreme and clear
  cases, regarding chaos). For energies more bounded the presence of
  chaos diminishes, but the general behavior is similar, i.e., if the
  pitch angle (or mass) increases chaos increases in the different
  energies. However, when chaos becomes pervasive and the main
  periodic orbits are destroyed, the chaotic behavior dominates in
  bounded energies as much as closer to corotation.

\begin{figure*}
\includegraphics[width=.95\textwidth]{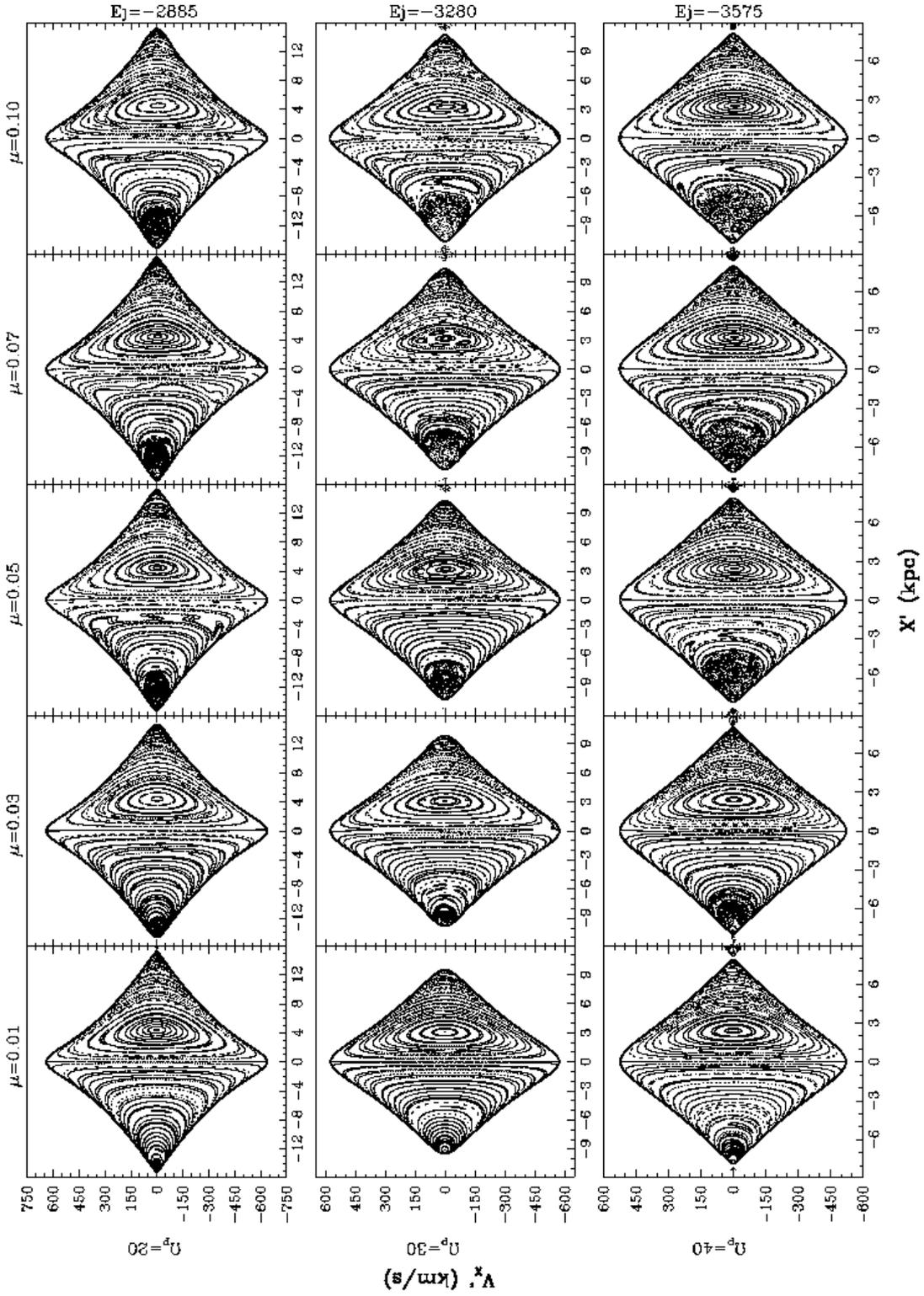}
\caption{Phase-space diagrams for an Sa galaxy with $i$ = $10\deg$.
  The values of the Jacobi energy, $\mu$, and $\Omega_{\rm p}$, are
  given at the right, top, and left, respectively.}
\label{DPmass_sa_pa10}
\end{figure*}

\begin{figure*}
\includegraphics[width=.95\textwidth]{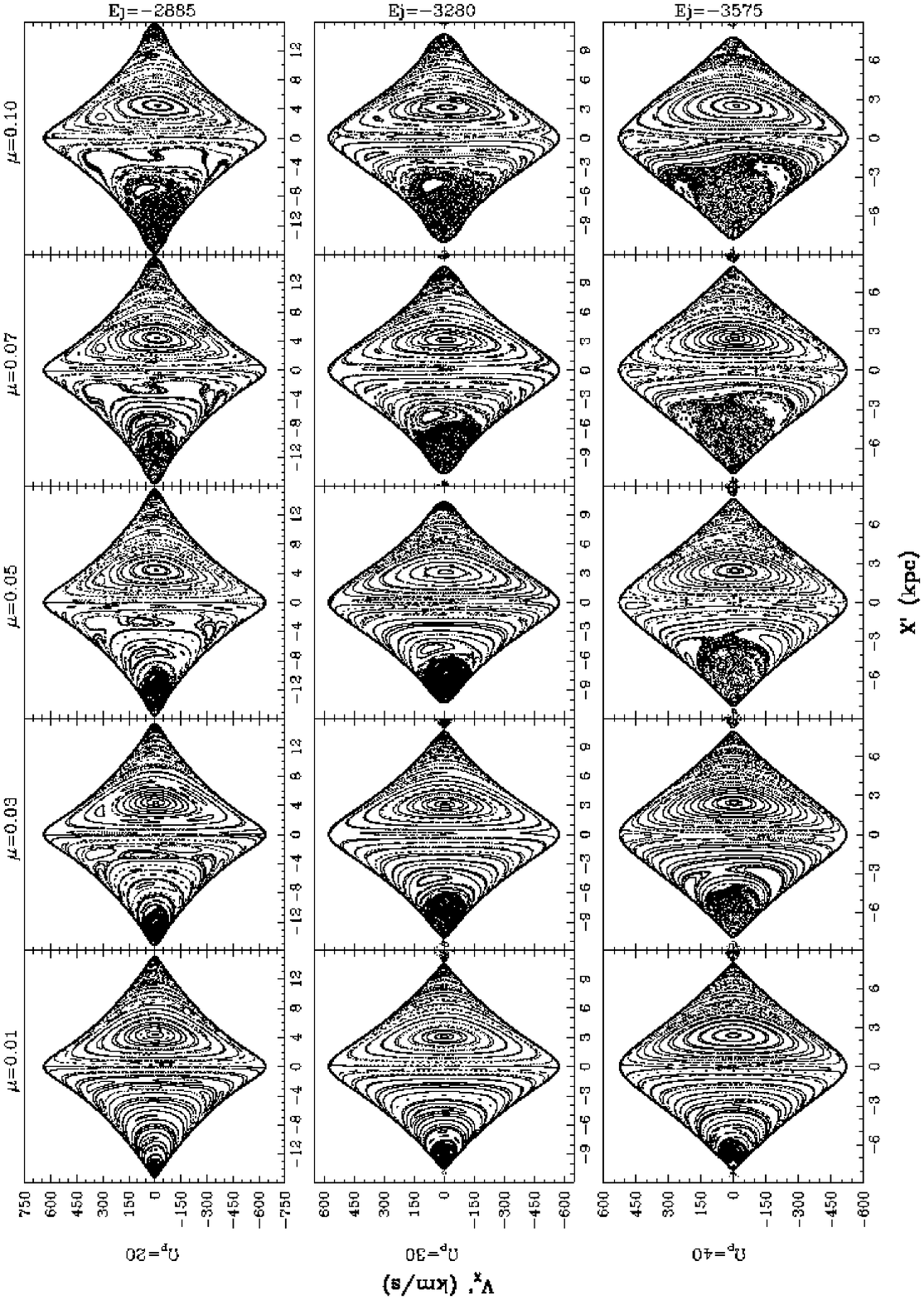}
\caption{As in Figure \ref{DPmass_sa_pa10}, here with $i$ = $20\deg$.}
\label{DPmass_sa_pa20}
\end{figure*}

\begin{figure*}
\includegraphics[width=.95\textwidth]{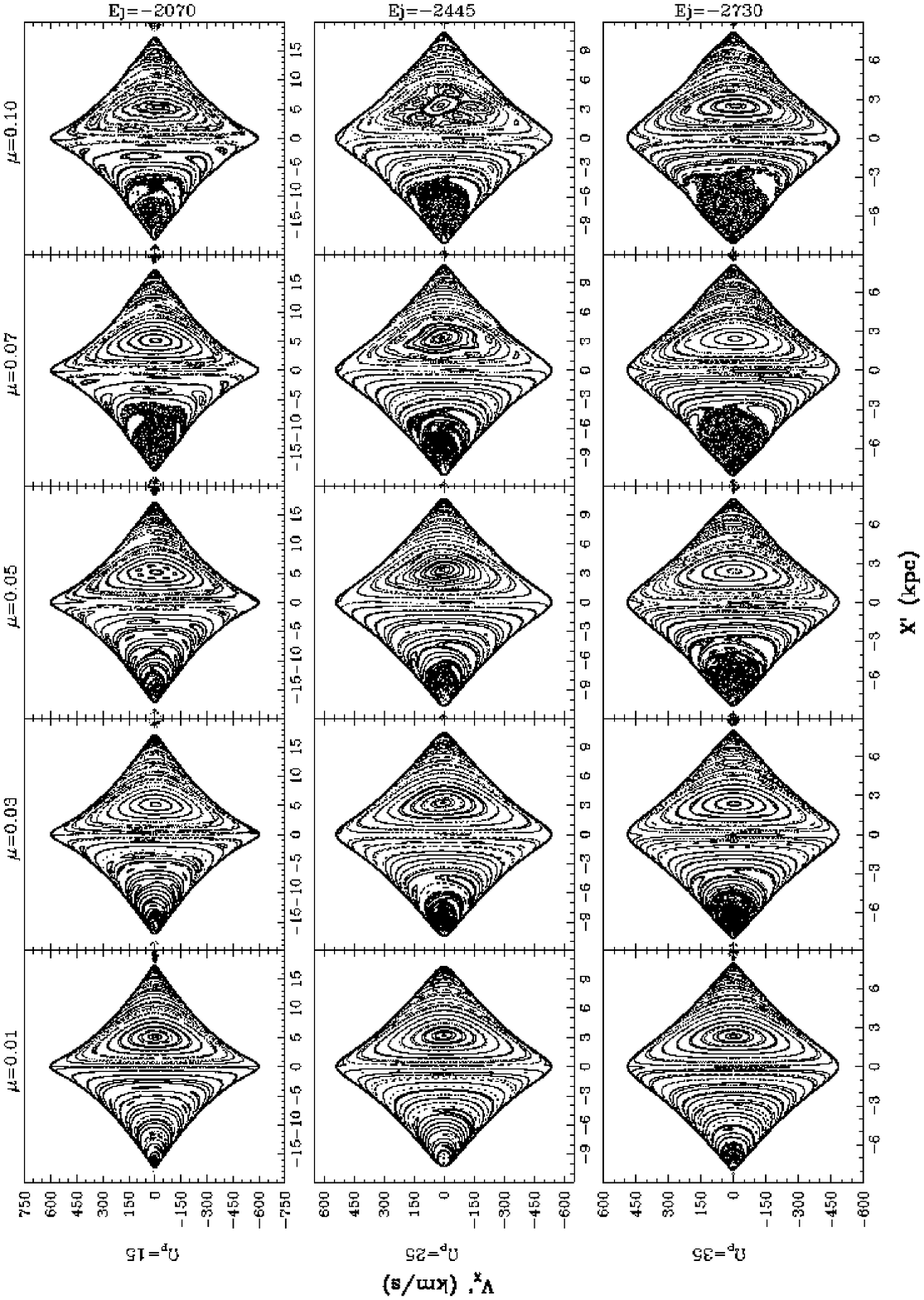}
\caption{Phase-space diagrams for an Sb galaxy with $i$ = $15\deg$.
  The values of the Jacobi energy, $\mu$, and $\Omega_{\rm p}$, are
  given at the right, top, and left, respectively.}
\label{DPmass_sb_pa15}
\end{figure*}

\begin{figure*}
\includegraphics[width=.95\textwidth]{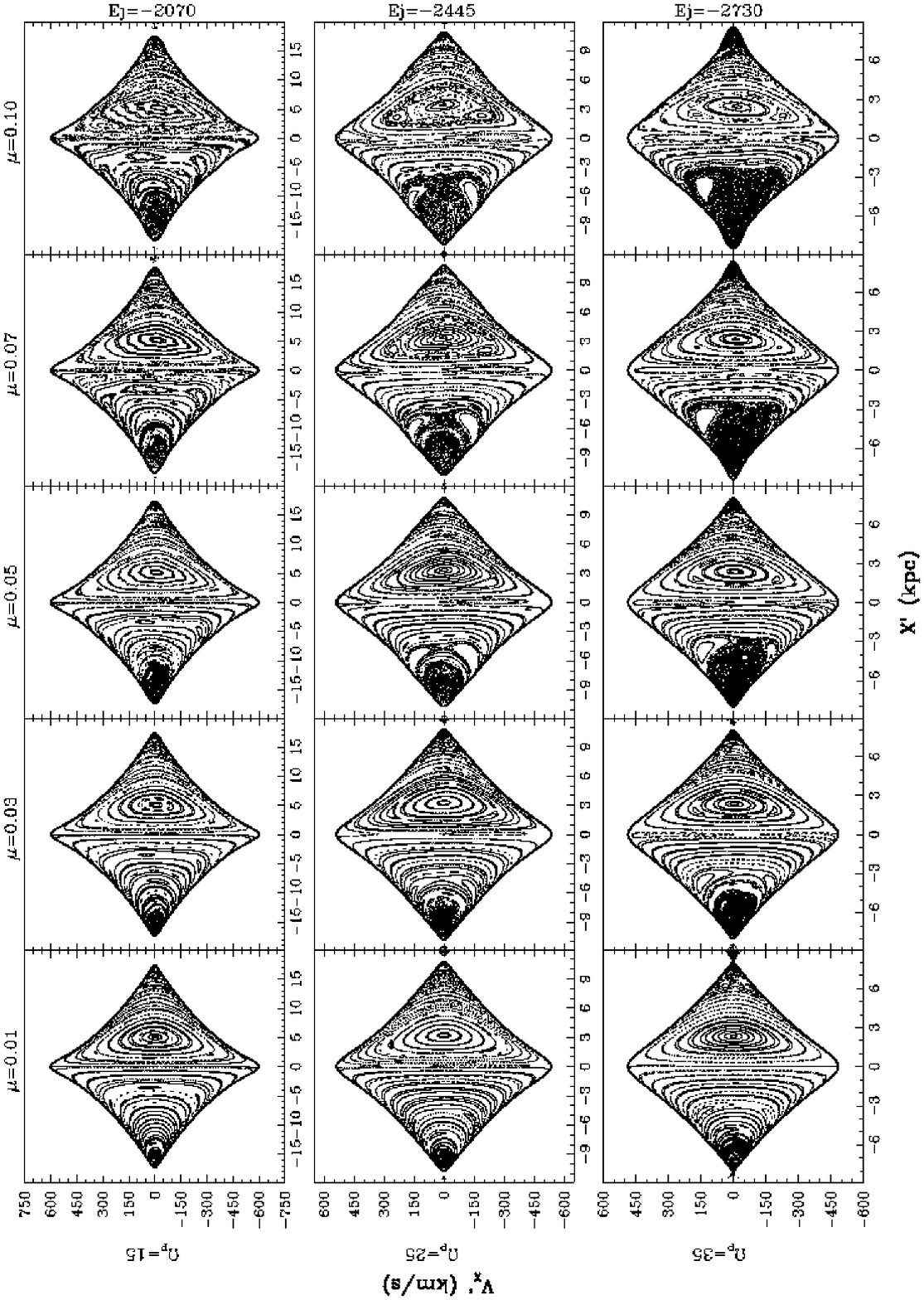}
\caption{As in Figure \ref{DPmass_sb_pa15}, here with $i$ = $20\deg$.}
\label{DPmass_sb_pa20}
\end{figure*}

\begin{figure*}
\includegraphics[width=.95\textwidth]{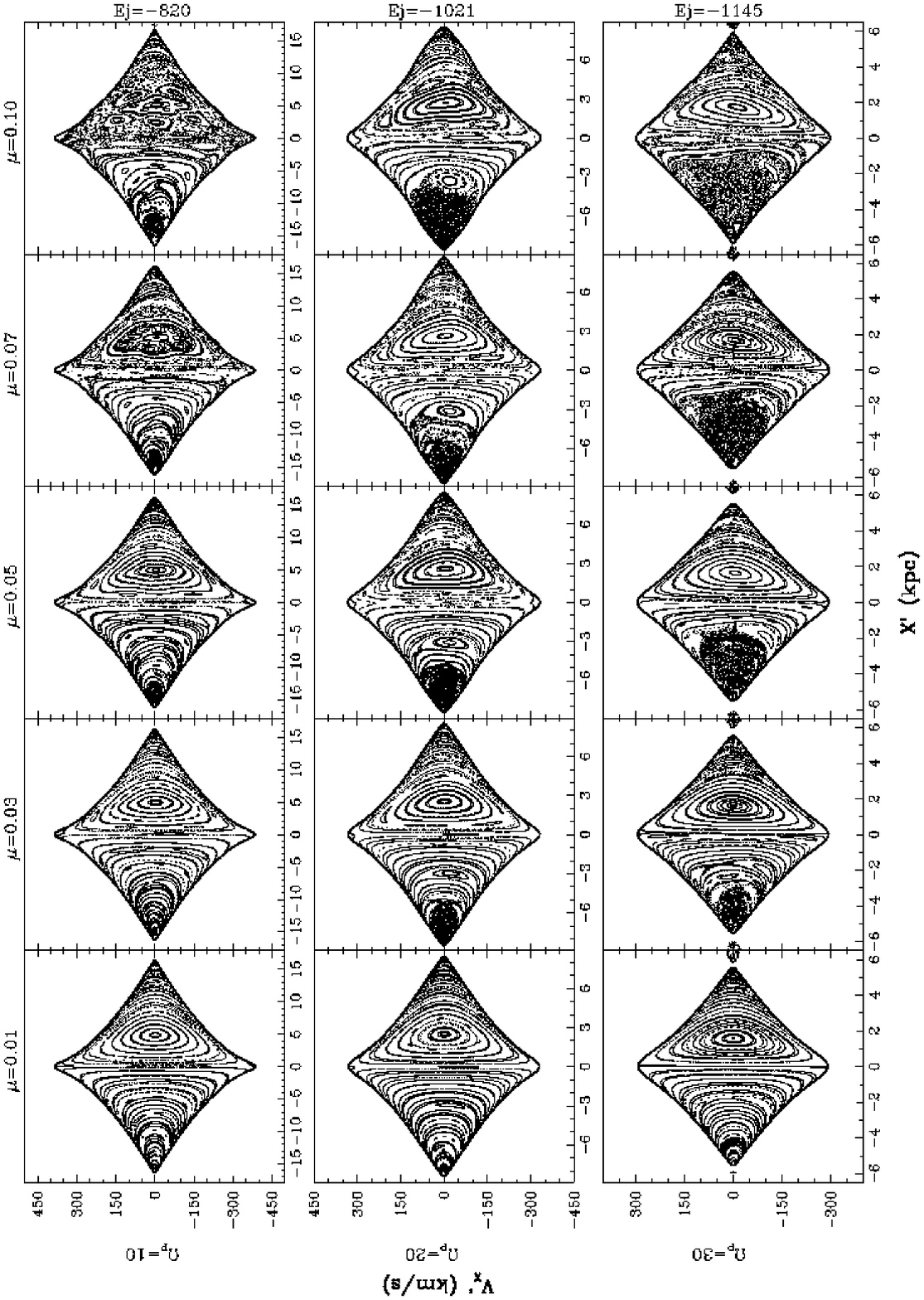}
\caption{Phase-space diagrams for an Sc galaxy with $i$ = $20\deg$.
  The values of the Jacobi energy, $\mu$, and $\Omega_{\rm p}$, are
  given at the right, top, and left, respectively.}
\label{DPmass_sc_pa20}
\end{figure*}

\begin{figure*}
\includegraphics[width=.95\textwidth]{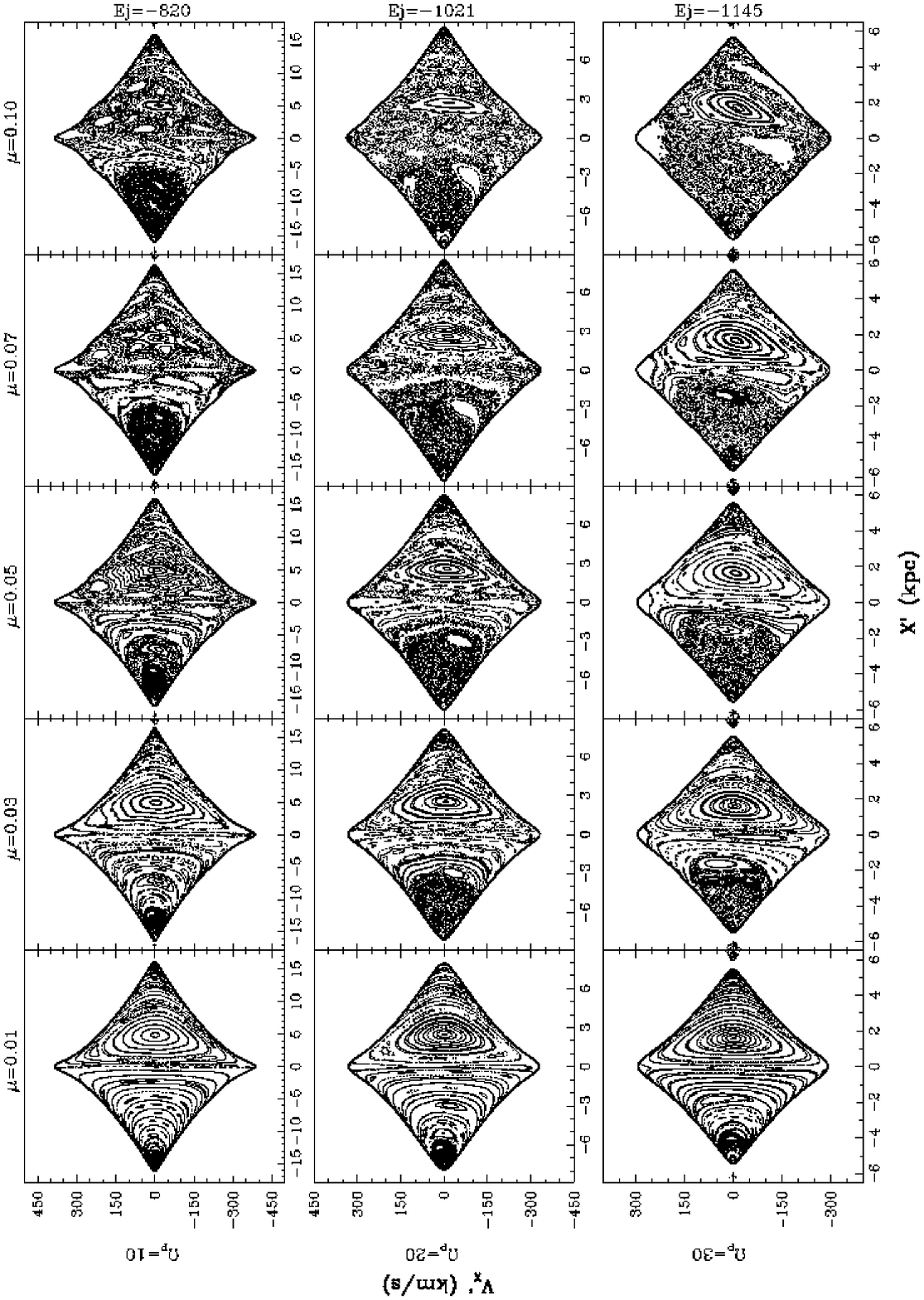}
\caption{As in Figure \ref{DPmass_sc_pa20}, here with $i$ = $30\deg$.}
\label{DPmass_sc_pa30}
\end{figure*}

Figures \ref{DPmass_sa_pa10} to \ref{DPmass_sc_pa30} show phase-space
diagrams for Sa, Sb, and Sc galaxies, considering in each type two
values of the pitch angle. The common trend in all these diagrams is
that the chaotic region which appears in the prograde (left) sides
increases as $\mu$ and $\Omega_{\rm p}$ increase. This chaotic region
extends toward the inner galactic region, destroying periodic orbits
that could support the spiral arms. In each galactic type the chaotic
region is more extended for the larger employed value of $i$, and it
is also markedly stronger for an Sc galaxy.

In summary, analyzing orbital self-consistency through periodic
orbits, we find that in order to produce long-lasting spiral arms, the
ratio $\mu$ in early spiral galaxies can be much larger than in late
spiral galaxies without compromising the stability of the
arms. Consequently, when the pitch angle is smaller, the limit for
$\mu$ can be considerably larger.  Approximately, the intervals in $i$
and $\mu$ to obtain long-lasting spiral arms in this scheme are the
following: for an Sa galaxy with $\Omega_{\rm p} \sim 30 \kmskpc$,
$i\lesssim10\deg$ and $\mu \lesssim 0.07$; for an Sb galaxy with
$\Omega_{\rm p} \sim 25 \kmskpc$, $i\lesssim15\deg$ and $\mu \lesssim
0.05$; and for an Sc galaxy with $\Omega_{\rm p} \sim 20 \kmskpc$,
$i\lesssim20\deg$ and $\mu \lesssim 0.03$. For greater values than
these, the spiral arms would be rather explained as transient
structures. The limits for $\mu$ are only examples that depend on the
values of $i$ and $\Omega_{\rm p}$; this means that these parameters
are deeply interrelated.

Regarding the chaotic behavior, with phase-space studies we also found
a maximum value for $\mu$, before chaos becomes pervasive destroying
the main periodic orbits which give support to spiral arms. As we
mentioned in the ordered case, the maximum limit of $\mu$ is mainly
linked to $i$ and less to $\Omega_{\rm p}$. Therefore, when $i$ is
smaller, the limit for $\mu$ can be larger, this is due to both
parameters are related to the spiral arm force (or amplitude of the
force). An example of the limits for $\mu$ depending on $i$ and
$\Omega_{\rm p}$ are: for an Sa galaxy, $\mu \lesssim 0.1$, with
$i\lesssim 20\deg$ and $\Omega_{\rm p} \sim 40 \kmskpc$; for an Sb
galaxy, $\mu \lesssim 0.07$, with $i\lesssim 20\deg$ and $\Omega_{\rm
  p} \sim 35 \kmskpc$; and for an Sc galaxy, $\mu \lesssim 0.05$, with
$i\lesssim 30\deg$ and $\Omega_{\rm p} \sim 20 \kmskpc$. For grater
values of $\mu$ the spiral arms are destroyed by chaotic behavior.

This analysis are some selected examples to clarify the general
orbital behavior. In Section \ref{valid_parameters} we will summarize
in a set of plots the ordered and chaotic behavior, taking a
significant increase in the number of values of the parameters $\mu$
and $i$ employed.

\subsection{Orbital Study Analyzing the Effect of the Angular Velocity
of the Spiral Arms: Ordered and Chaotic Behavior}\label{angular_speed}

In Section \ref{mass}, we have analyzed the effect of the mass of the
spiral arms on the ordered and chaotic stellar dynamics on the
equatorial plane of normal spiral galaxies; for this purpose we
employed assorted masses, pitch angles and angular velocities of the
spiral arms. In this Section we present a similar orbital study,
analyzing the effect in the ordered and chaotic stellar dynamics as we
vary $\Omega_{\rm p}$ in an extended interval, from 10 to 60 $\kmskpc$
for each morphological type. As in the case of the spiral-arms-mass
analysis, in order to dilucidate their relative importance, we also
slightly change other parameters; for the pitch angle we take
respectively the values $7\deg$, $18\deg$, and $25\deg$, in Sa, Sb, Sc
galactic types, and in all these galactic types $\mu$ takes the values
0.01, 0.03, and 0.05. As in the previous subsection, these are only
some examples to obtain a perception of the dynamical behavior
exerted by changes in the spiral arms parameters. In the next section
we summarize the results.

\begin{figure*}
\includegraphics[width=1\textwidth]{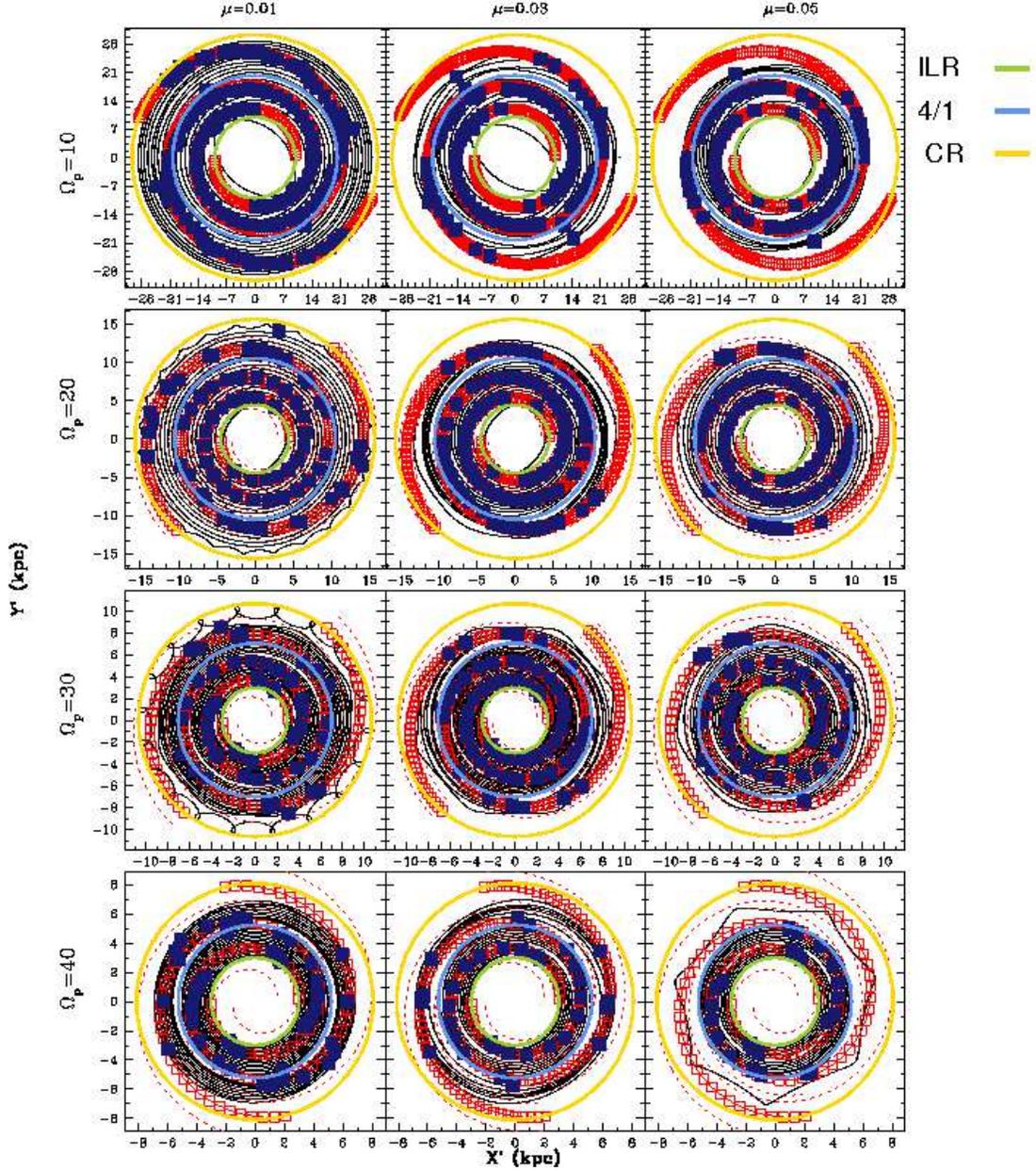}
\caption{\small Periodic orbits (black curves), density response
  maxima (filled squares), and the imposed spiral arms locus (open
  squares and dotted lines mark the width of spiral arms) for the
  three-dimensional spiral arms model of an Sa galaxy with a pitch
  angle $i$ = $7\deg$. The values of $\mu$ and $\Omega_{\rm p}$ are
  given at the top and left, respectively.}
\label{p_omega_sa_pa7}
\end{figure*}

In Figure \ref{p_omega_sa_pa7} we present periodic orbits for an Sa
galaxy with $i$ = $7\deg$. This figure shows that the amount of
periodic orbits which give support to the spiral arms decrease with
$\Omega_{\rm p}$ and $\mu$. For $\Omega_{\rm p}\lesssim 30 \kmskpc$,
the density response follows the imposed spiral arms potential almost
to the CR position; for $\Omega_{\rm p}\sim 40 \kmskpc$, the density
support extends slightly beyond the 4/1 resonance position. This
behavior is obtained with $\mu \sim 0.01$. If $\mu$ increases between
0.03 and 0.05 the density support extends beyond the 4/1 resonance
position. For $\Omega_{\rm p}>40 \kmskpc$, there is no density
support.  If we increase the pitch angle to $18\deg$, the density
support extends almost to the CR position only if $\mu$ $\simless$
0.01; above this value of $\mu$ there is a density support up to the
4/1 resonance. If $\Omega_{\rm p} >30 \kmskpc$ and $\mu$ $\simless$
0.05, the density support reaches the 4/1 resonance position. Thus,
the orbital support depends on three parameters of the spiral arms:
their pitch angle, mass, and angular speed, being the dependence on
the first parameter, the more sensitive. The value of this parameter
is the one with the wider possible range in galaxies.

\begin{figure*}
\includegraphics[width=1\textwidth]{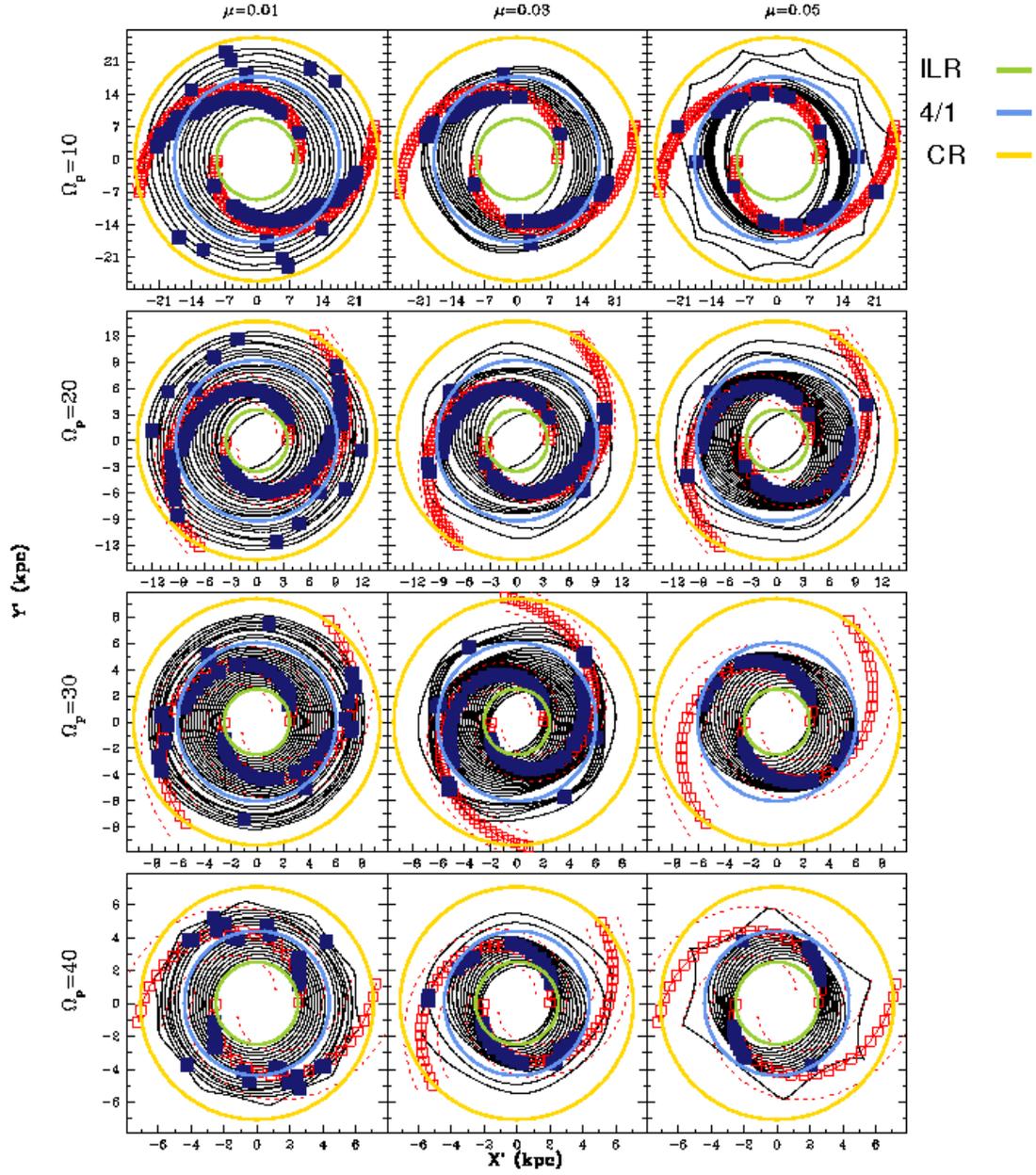}
\caption{\small As in Figure \ref{p_omega_sa_pa7}, here for an Sb
  galaxy with $i$ = $18\deg$.} 
\label{p_omega_sb_pa18}
\end{figure*}

Figure \ref{p_omega_sb_pa18} shows periodic orbits for an Sb galaxy
with $i$ = $18\deg$. For $\Omega_{\rm p}\lesssim30 \kmskpc$ and
$\mu$ = 0.01, the density support extends approximately up to the CR 
position. For $\Omega_{\rm p}=40 \kmskpc$ and $\mu$ = 0.01 this
support extends only up to the 4/1 resonance. 
If $\mu$ $>$ 0.01 the density support extends up to the 4/1 resonance
but producing a pitch angle smaller than in the imposed arms. For cases
where $\Omega_{\rm p} >40 \kmskpc$, there is no density support.

\begin{figure*}
\includegraphics[width=1\textwidth]{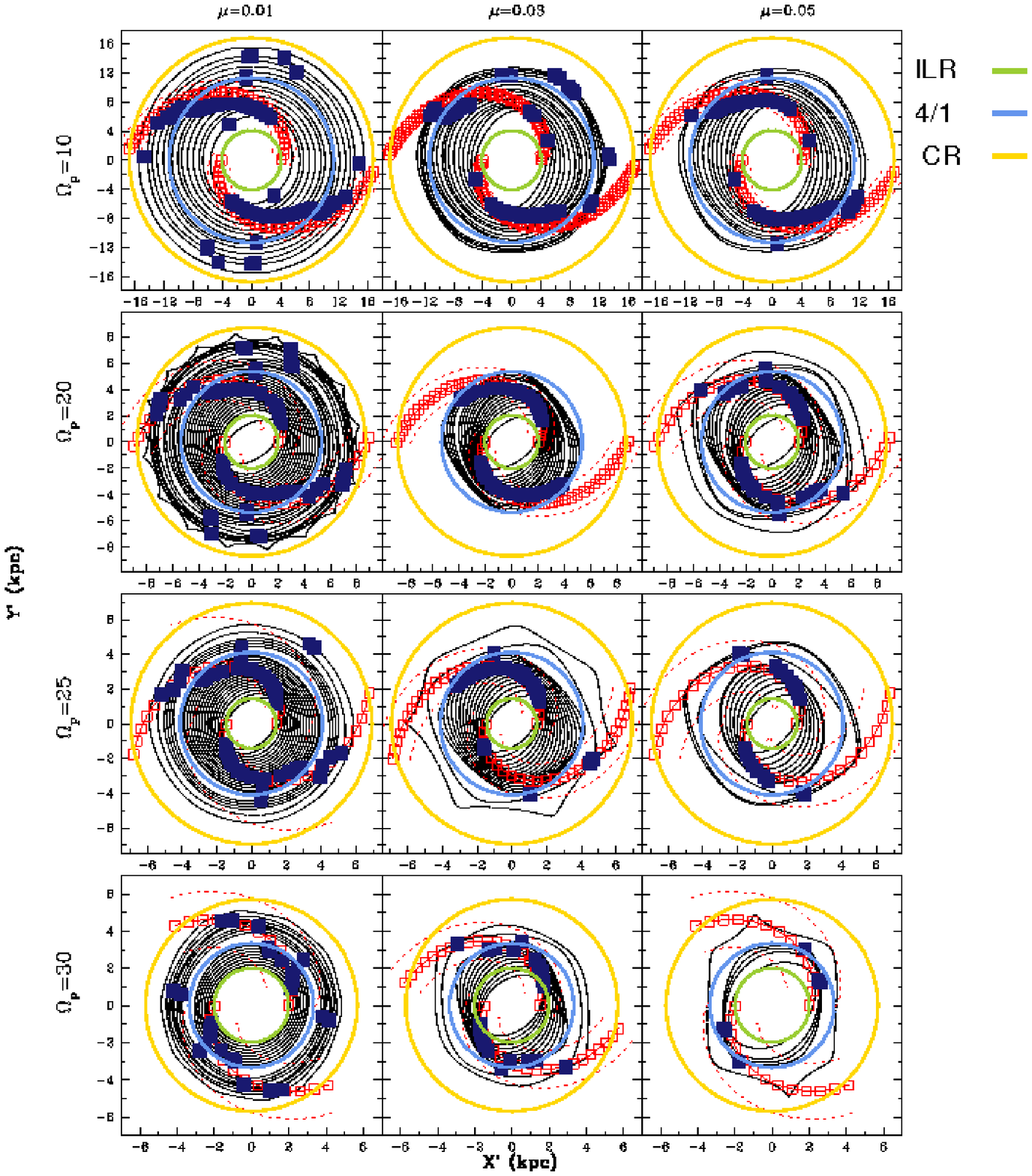}
\caption{\small As in Figure \ref{p_omega_sa_pa7}, here for an Sc 
  galaxy with $i$ = $25\deg$.} 
\label{p_omega_sc_pa25}
\end{figure*}

In Figure \ref{p_omega_sc_pa25} we show periodic orbits for an Sc
galaxy with $i$ = $25\deg$. For $\Omega_{\rm p}\lesssim30$ and
$\mu$ = 0.01, the density support extends not far from the
CR position. For $\mu$ = 0.03, 0.05, this support extends 
up to the 4/1 resonance position (in some cases slightly beyond) 
forming a smaller pitch angle than in the imposed arms. 
With $\mu$ = 0.05 there is no density support if   
$\Omega_{\rm p}\sim30 \kmskpc$ or larger.

\begin{figure*}
\includegraphics[width=1\textwidth]{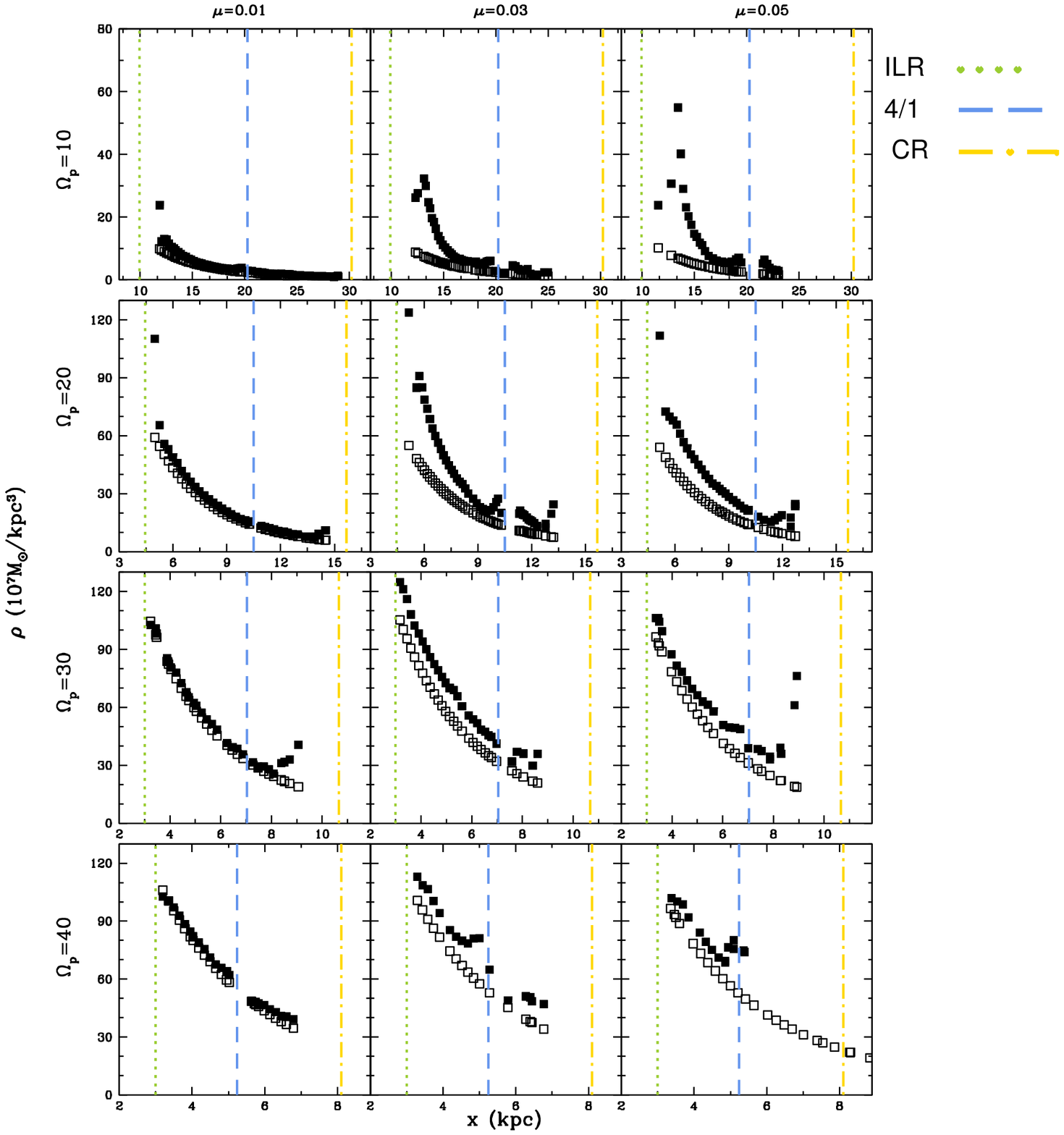}
\caption{ \small Density response diagrams. Filled squares are the
  density response of spiral arms for an Sa galaxy, and open squares
  represent the imposed density with a pitch angle $i$ = $7\deg$. The
  values of $\mu$ and $\Omega_{\rm p}$ are given at the top and left,
  respectively. The dotted, dashed and dot-dashed lines show the ILR
  position, 4/1 resonance position and CR position, respectively.}
\label{r_omega_sa_pa7}
\end{figure*}

\begin{figure*}
\includegraphics[width=1\textwidth]{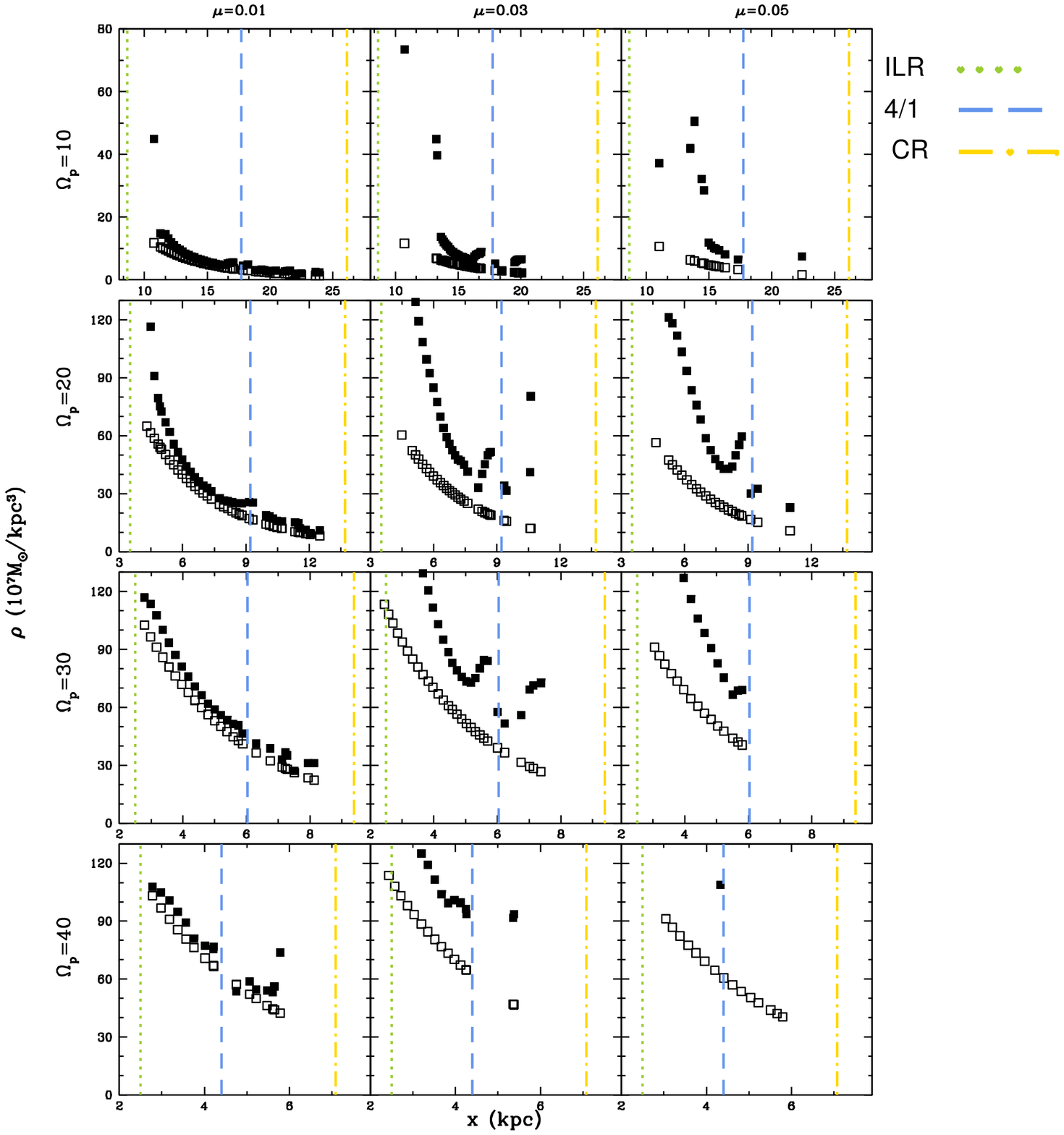}
\caption{\small As in Figure \ref{r_omega_sa_pa7}, here for an Sb
  galaxy with $i$ = $18\deg$.} 
\label{r_omega_sb_pa18}
\end{figure*}

\begin{figure*}
\includegraphics[width=1\textwidth]{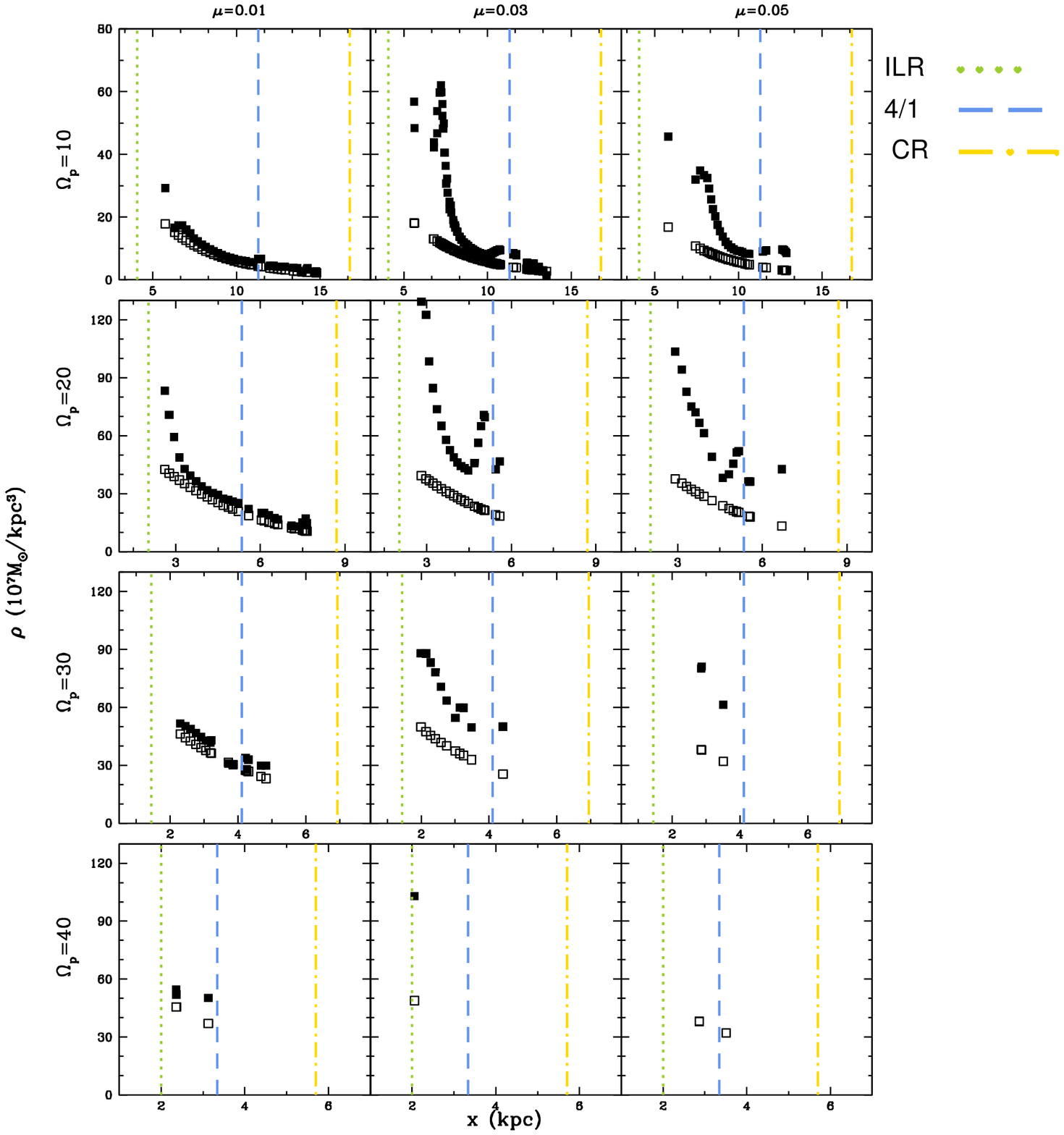}
\caption{\small As in Figure \ref{r_omega_sa_pa7}, here for an Sc 
  galaxy with $i$ = $25\deg$.} 
\label{r_omega_sc_pa25}
\end{figure*}

 As we did in the case where we analyzed the effect of spiral arms
  mass, we have compared the spiral arms density response (filled
  squares) with the spiral arms imposed density (PERLAS, open
  squares). We constructed a mosaic of density response corresponding
  to each periodic orbit mosaic. Figures \ref{r_omega_sa_pa7},
  \ref{r_omega_sb_pa18} and \ref{r_omega_sc_pa25} show the densities
  for Sa, Sb and Sc galaxies, respectively. With these mosaics we
  reinforce the results presented with periodic orbits and maxima
  density response in Figures \ref{p_omega_sa_pa7} -
  \ref{p_omega_sc_pa25}.

Now, in order to analyze the chaotic behavior, as was done varying the mass
of spiral arms (Section \ref{mass}), in this part we also present a
detailed study of Poincar\'e diagrams varying the angular speed of the
spiral arms in an extended interval. We change slightly the mass of
the spiral arms and the pitch angle. With this analysis we found a
limit to the angular speed for which chaos becomes pervasive,
dominating the available phase space and destroying all the periodic
orbits as well as the ordered orbits surrounding them.

\begin{figure*}
\includegraphics[width=1\textwidth]{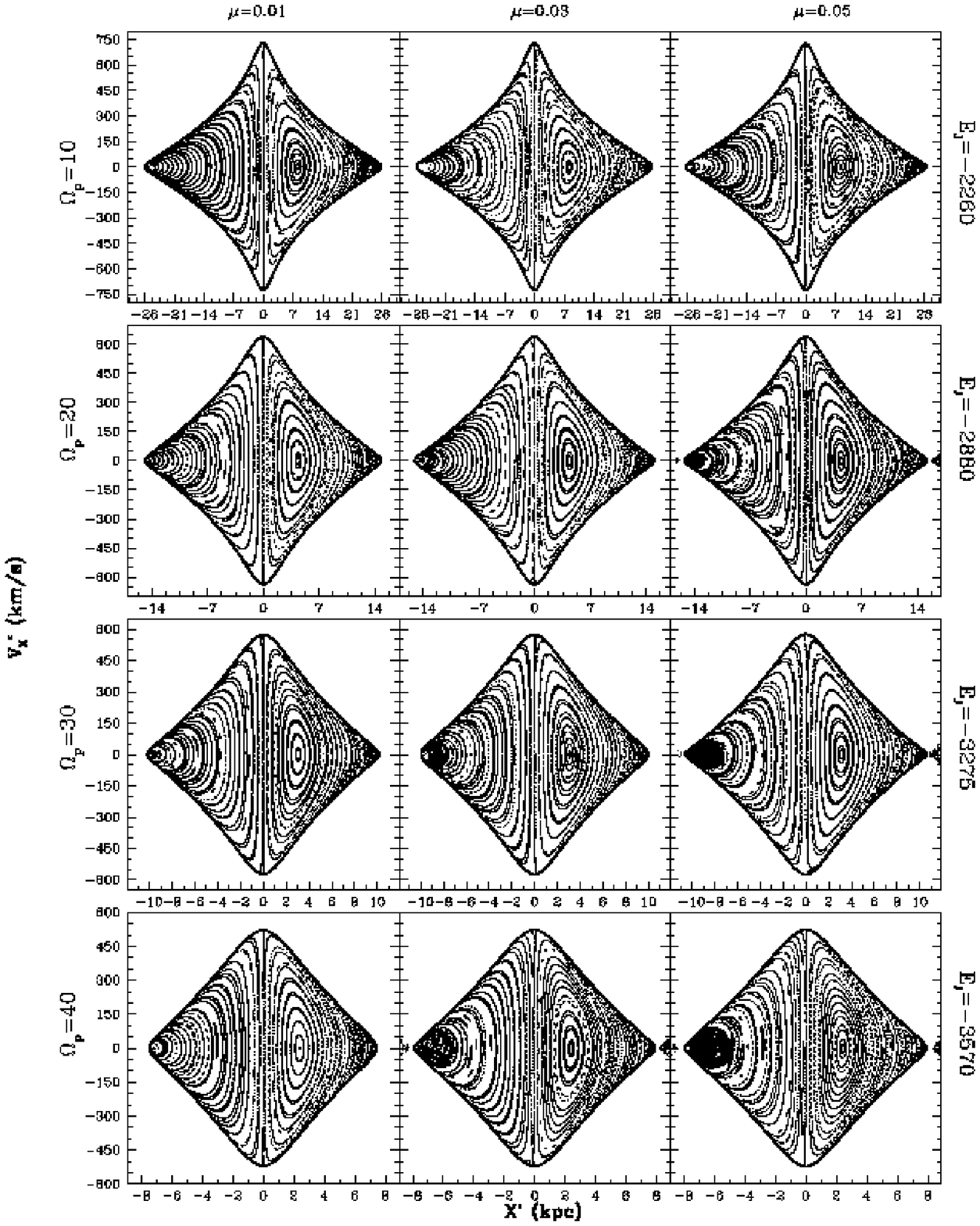}
\caption{Phase-space diagrams for an Sa galaxy with $i$ = $7\deg$.
  Here a more extended interval of values for $\Omega_{\rm p}$ is
  considered, compared with that employed in Section \ref{mass}. 
  The values of the Jacobi energy, $\mu$, and $\Omega_{\rm p}$, are
  given at the right, top, and left, respectively.}
\label{DPomega_sa_pa7}
\end{figure*}

Figure \ref{DPomega_sa_pa7} shows Poincar\'e diagrams for an Sa galaxy
with $i$ = $7\deg$. For $\Omega_{\rm p}=10$ and $20 \kmskpc$ the
orbital behavior is ordered; however the chaotic regions emerge when
$\mu$ increases from 0.01 to 0.05. For $\Omega_{\rm p}=30$ and
$40 \kmskpc$, the ordered orbits dominate, but even with $\mu=0.01$ the
chaotic behavior already appears in the prograde region, and increases
with $\mu$. If $i$ = $15\deg$ the chaotic behavior increases with the
mass and angular speed of the spiral arms.

\begin{figure*}
\includegraphics[width=1\textwidth]{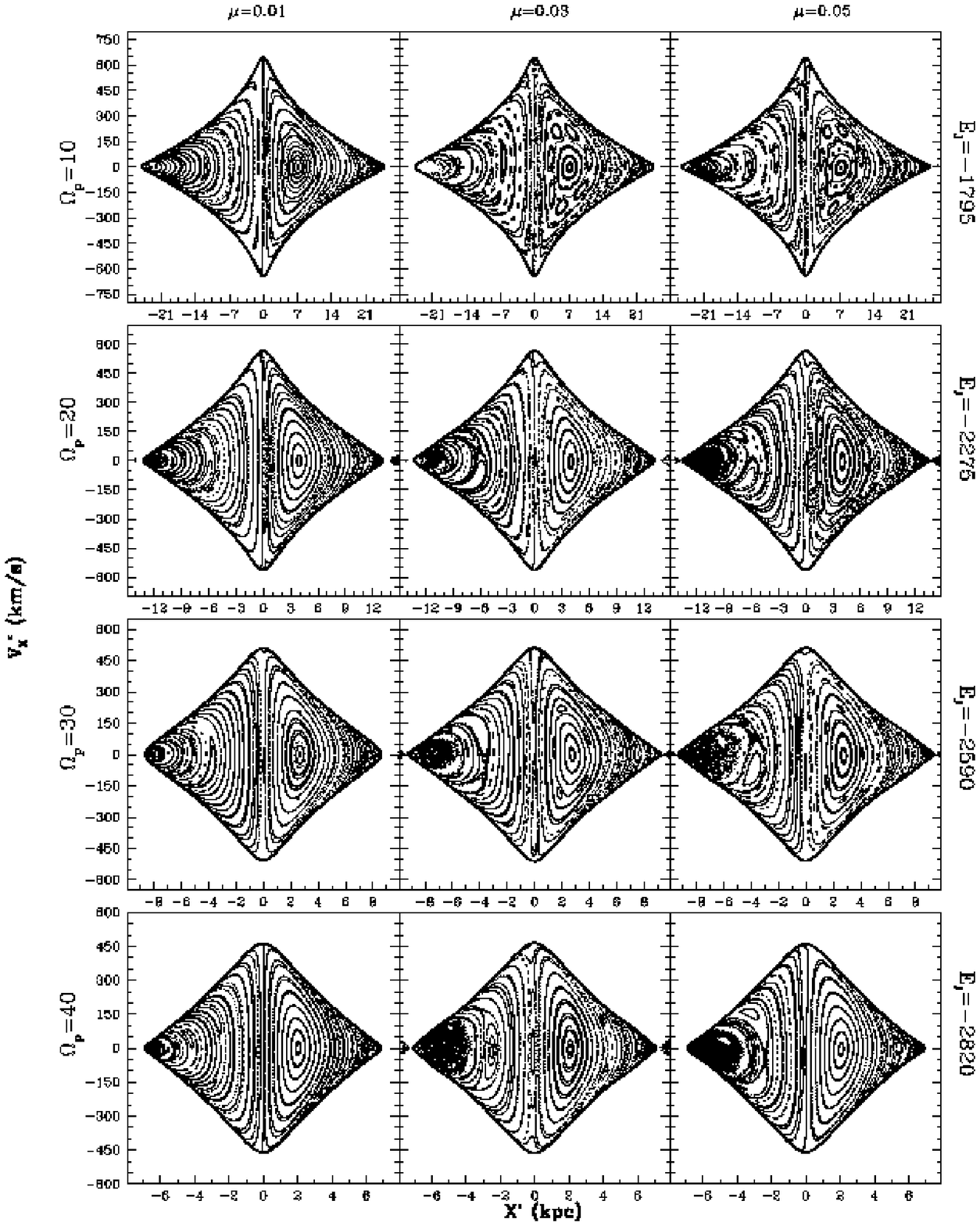}
\caption{As in Figure \ref{DPomega_sa_pa7}, here for an Sb galaxy
with $i$ = $18\deg$.}  
\label{DPomega_sb_pa18}
\end{figure*}

Figure \ref{DPomega_sb_pa18} shows Poincar\'e diagrams for an Sb galaxy
with $i$ = $18\deg$. The ordered orbits dominate the prograde region
and there is a small region of chaos, which increases slightly with
$\Omega_{\rm p}$. Additionally, the orbits are more complex and
resonant islands appear. A severe increment of the chaotic region towards the main periodic orbits supporting the spiral arms is related
with an increment of $\mu$; for example, when $\mu$ = 0.05 and
$\Omega_{\rm p}=40\kmskpc$, the chaotic behavior covers an important
part of the prograde region.

\begin{figure*}
\includegraphics[width=1\textwidth]{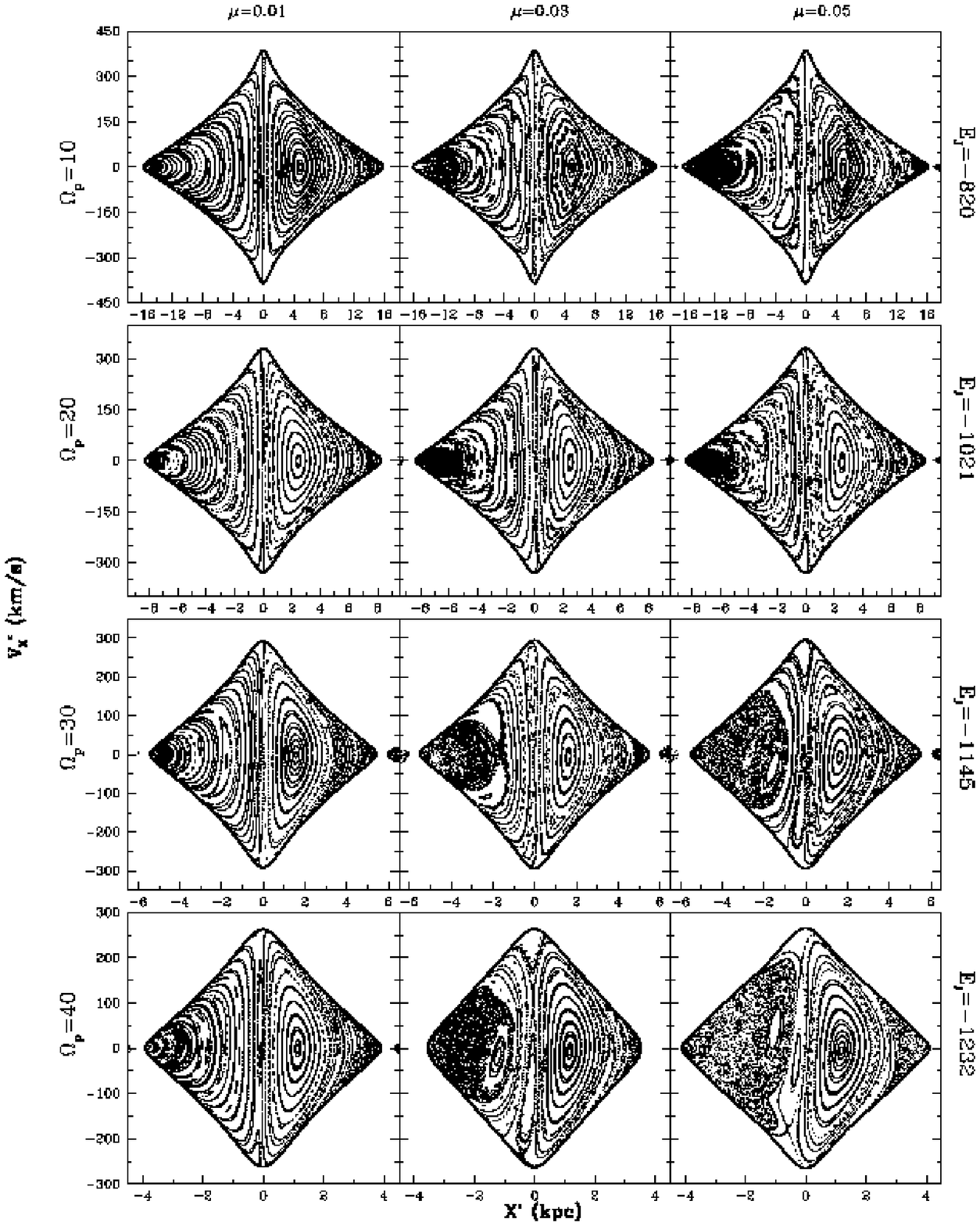}
\caption{As in Figure \ref{DPomega_sa_pa7}, here for an Sc galaxy 
with $i$ = $25\deg$.}                   
\label{DPomega_sc_pa25}
\end{figure*}

Figure \ref{DPomega_sc_pa25} shows Poincar\'e diagrams for an Sc
galaxy with $i$ = $25\deg$. The majority of orbits are ordered, but
the chaotic region slowly increases with $\Omega_{\rm p}$. The chaotic
behavior is more prone to emerge when $\mu$ is larger; for example, if
$\mu$ = 0.05 and $\Omega_{\rm p} = 40 \kmskpc $ chaos dominates an
important region of the available phase-space, covering practically
all the prograde region and destroying the periodic orbits.

In summary, regarding ordered behavior, we constrain the angular speed
of the spiral arms through the existence of periodic orbits. We found
that the orbital support for these arms depends on three parameters:
the pitch angle, mass, and angular speed, but the orbital support
seems to be much more sensitive first to the pitch angle, and second
to the mass of spiral arms, and almost insensitive to the angular
speed (although it is important because this defines the extension of
the spiral arms). With our analysis, we set a limit to $\Omega_{\rm
  p}$ for each morphological type taking into account the three
parametes: for an Sa galaxy, $i\lesssim7\deg$, $\Omega_{\rm p} \sim 40
\kmskpc$, and $\mu \lesssim 0.03$; for an Sb galaxy,
$i\lesssim18\deg$, $\Omega_{\rm p} \sim 30\kmskpc$, and $\mu \lesssim
0.03$; and for an Sc galaxy, $i\lesssim25\deg$, $\Omega_{\rm p} \sim
25 \kmskpc$, and $\mu \lesssim 0.01$. For laeger values, there are not
enough periodic orbits to provide support to the spiral arms. These
limits for $i$, $\Omega_{\rm p}$, and $\mu$ are only examples. In the
following section we provide a general analysis.

Regarding chaotic behavior, with Poincar\'e diagrams we also found a
maximum value for $\mu$, before chaos becomes pervasive destroying all
the main periodic orbits which give support to spiral arms. As we
found for the ordered case, the maximum limit of $\Omega_{\rm p}$ is
linked to $i$ and $\mu$.  An example of the limits for $\Omega_{\rm
  p}$ depending on $i$ with $\mu=0.05 $ are: for an Sa galaxy with
$i\lesssim 7\deg$, $\Omega_{\rm p} \sim 40 \kmskpc$; for an Sb galaxy
with $i\lesssim 18\deg$, $\Omega_{\rm p} \sim 40 \kmskpc$; and for an
Sc galaxy with $i\lesssim 25\deg$, $\Omega_{\rm p} \sim 30 \kmskpc$.

\section {Limits to Parameters for Plausible Dynamical Models for Spiral Arms} \label{valid_parameters}

Considering the analysis of the effect of the pitch angle that was
performed in Paper I, and with the examples obtained in Section
\ref{results} concerning the effects of the mass and angular speed of
the spiral arms, in this section we present an extended analysis
increasing the number of values studied of the parameters $i$ and
$\mu$ in normal spiral galaxies. We study the ordered and chaotic
behavior on the galactic plane, through periodic orbits, maxima
density response and Poincar\'e diagrams.  We present two maximum
limits for these parameters. The first of them is regarding periodic
orbits and density response, where the imposed spiral arms are
supported by the density response; this could tell us about the nature
of spiral arms: if they are transient or long-lasting. The second
limit is a detailed analysis based on phase space diagrams (regardless
the spiral arms nature), before the chaotic behavior becomes pervasive
dominating all available phase space, and destroying all orbital
support.

In Paper I and in P\'erez-Villegas \et (2012) we presented a deep
study of the effects of the pitch angle on normal spiral galaxies. The
purpose of that study was to provide the values of the pitch angle in
galaxies that produced transient or long-lasting galactic models
(assuming typical masses for the axisymmetric background potential for
disk galaxies from Sa to Sc morphological types). In those studies we
considered only the effect of the pitch angle, keeping fixed the mass
and angular speed of the spiral arms.

\begin{figure*}
\includegraphics[width=.95\textwidth]{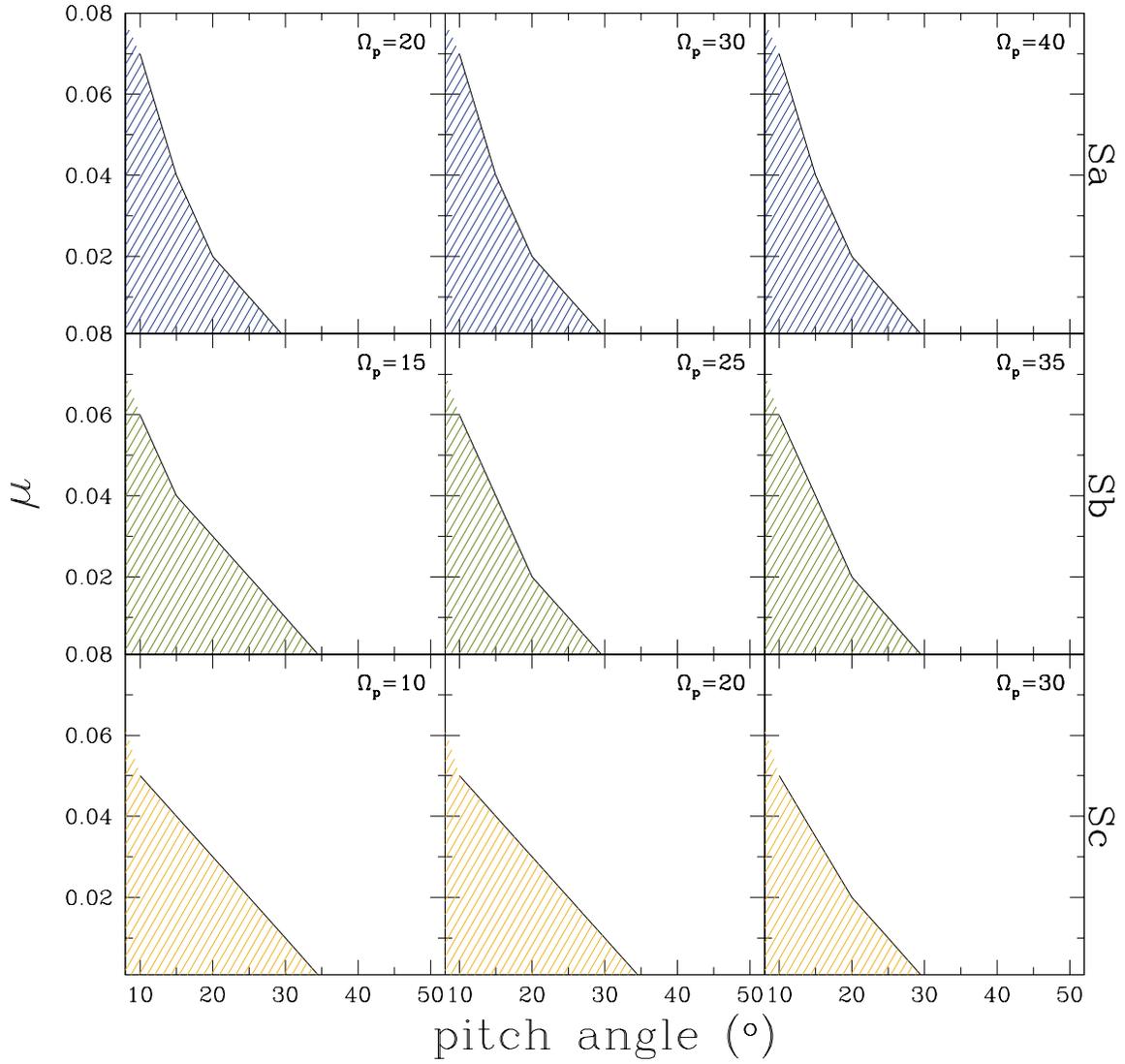}
\caption{Dynamically plausible models for Sa (top line), Sb (middle
  line), and Sc (bottom line) galaxies. The shaded regions provide the
  parameters to construct long-lasting spiral arms models in the
  scheme presented in this work. Spiral arms with parameters outside
  the shaded regions, would most likely act as transient features.}
\label{mod_auto_orden}
\end{figure*}

In Figure \ref{mod_auto_orden} we summarize our results. First we
present a $3\times3$ mosaic, that shows the maxima values of $\mu$ to
obtain long-lasting spiral arms, in Sa, Sb, Sc galaxies, depending on
the pitch angle $i$ and angular speed $\Omega_{\rm p}$. Beyond these
values the spiral arms would be considered as transient features.

Based in the detailed phase-space orbital study presented in
P\'erez-Villegas \et (2012) and Paper I, concerning the restriction of
the pitch angle given by the chaotic behavior in normal spiral
galaxies, here we present maximum values for structural and dynamical
parameters of spiral arms such as pitch angle, mass, and angular
speed, before the chaotic behavior dominates the available phase-space
destroying the main stable periodic orbits. Large-scale structures
such as spiral arms are not expected to appear in galaxies where
chaotic behavior dominates completely (Voglis \et 2006); however,
confined chaotic orbits may provide some support to spiral arms
(Patsis \& Kalapotharakos 2011; Kaufmann \& Contopoulos 1996;
Contopoulos \& Grosb\o l 1986), up to a certain point, prior to the
destruction of all phase space surrounding the periodic orbits that
give the shape to spiral arms.

\begin{figure*}
\includegraphics[width=.95\textwidth]{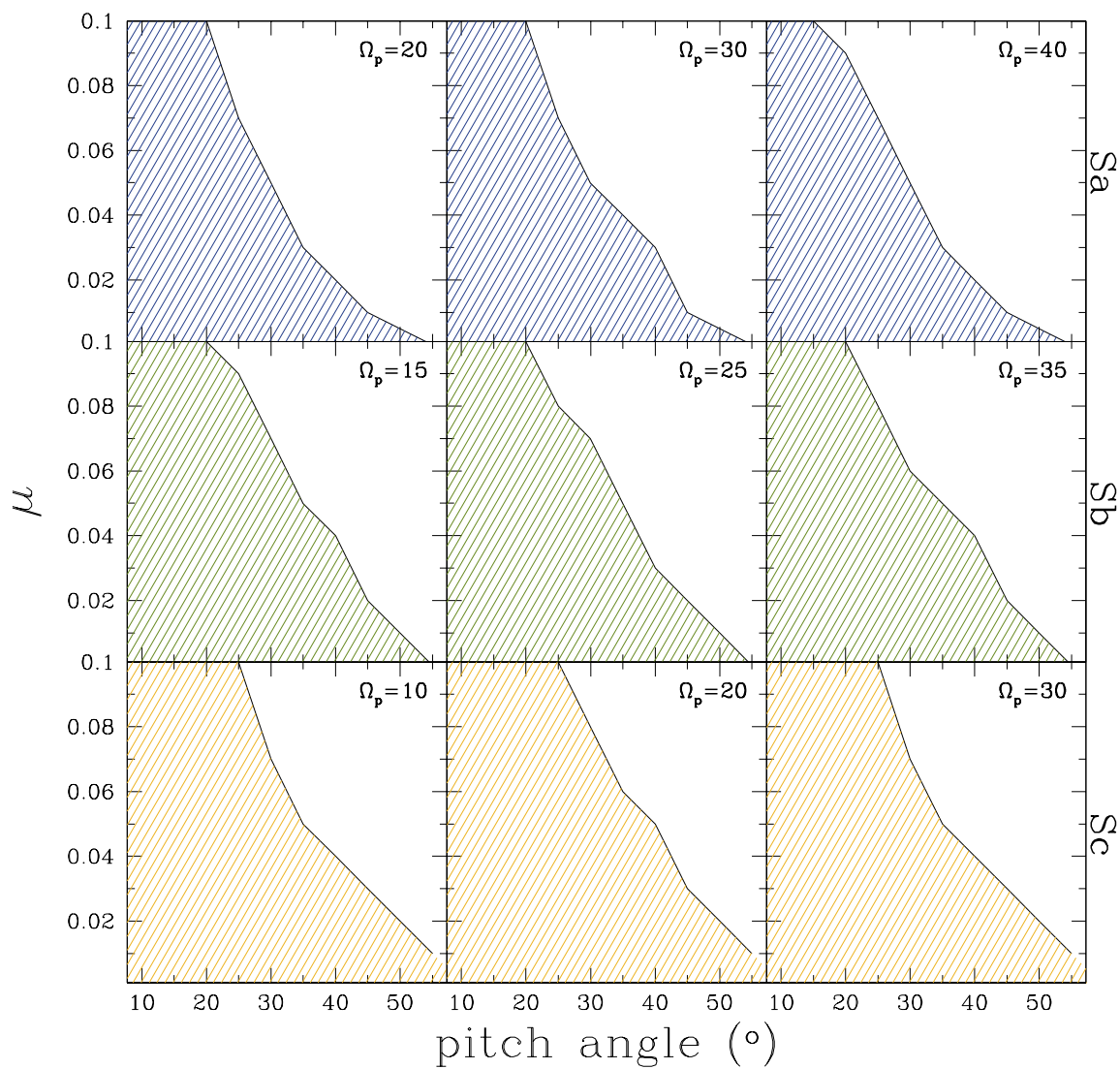}
\caption{Models for Sa (top line), Sb (middle line), and Sc (bottom
  line) galaxies. The solid line is the maximum limit for the spiral
  arms models before the domain of chaotic behavior. Parameters for
  spiral arms on the shaded regions, would be dynamically plausible
  (independently of their likely transient nature).}
\label{mod_auto_caos}
\end{figure*}

In Figure \ref{mod_auto_caos} we present a $3\times3$ mosaic; each
panel shows permitted values of $\mu$ and pitch angle $i$, depending
on morphological type and angular speed, before chaos dominates.  For
values of the spiral arm parameters smaller than those given by the
continuous lines, the chaotic behavior may be important, but it is
still confined by stable quasiperiodic orbits. For values lager than
these limits, the chaotic behavior becomes pervasive destroying all
the available prograde phase-space.

\section{Discussion and Conclusions}\label{discussion} 

With the use of a family of models observationally motivated to
simulate typical Sa, Sb and Sc spiral galaxies, that includes a
bisymmetric density-based spiral arms potential model, we perform an
extensive analysis of the stellar dynamical effects of spiral arms on
galactic disks. The spiral arms model is a self-gravitating
  three-dimensional potential constructed with individual oblate
  spheroids (as bricks in a building); this means that the model
  produces a density based force field (i.e. a more physical model,
  instead of an {\it ad hoc} mathematical fit), which means in turn
  that the potential responds for example to changes on the structure
  such as a larger pitch angle that naturally produces that disk
  particles feel a more aggressive effect (i.e. the attack angle for
  particles is larger the larger the pitch angle is). 

In this work, we have extended the studies of Paper I (devoted only to
pitch angle effects), to the effects of the spiral arm strength (mass
of the spiral arms), its angular speed and pitch angle all
together. For all morphological types, we varied the mass of the
spiral arms within approximately 1 to 10\% of the mass of the disk,
and its angular speed from 10 to $60\kmskpc$. In Sa, Sb and Sc
galaxies, pitch angle employed values were $7\deg$ to $20\deg$,
$10\deg$ to $20\deg$, and $15\deg$ to $30\deg$, respectively.

As in paper I, we present two sets of restrictions different in nature
for spiral arms parameters. One is based on ordered dynamical behavior
and the second on chaotic behavior. Restrictions based on ordered
behavior, provide us a tool based on orbital support for the spiral
arms that refers to their transient vs. long-lasting nature. The
second set of restrictions, based on chaotic behavior, represents the
limits beyond which, spiral arms are no longer feasible.

For the first limit, we produced an orbital study based on periodic
orbits and computed the maxima density response comparing it with the
imposed potential to produce a set of plausible models for spiral
galaxies with more probable long-lasting spiral arms. In this case we
find that the mass of the spiral arms, M$_{\rm sp}$, should decrease
with the increase of the pitch angle $i$; if $i$ is smaller than
$\sim10\deg$, M$_{\rm sp}$ can be as large as $\sim10\%$, $\sim7\%$,
$\sim5\%$ of the disk mass, for Sa, Sb, and Sc galaxies,
respectively. If $i$ increases to $\sim25\deg$, M$_{\rm sp}$ is around
1\% of the mass of the disk for all morphological types. For values
larger than these limits, spiral arms would be transient features.

For the second limit, we produced a phase-space study with Poincar\'e
diagrams, based on chaotic orbital behavior.  We seek the parameters
of the spiral arms prior to the domain of chaos that destroys all
orbital support for the arms. In this case we also found that M$_{\rm
  sp}$ should decrease with the pitch angle $i$.  If $i$ is smaller
than $\sim20\deg$, $\sim25\deg$, and $\sim30\deg$, for Sa, Sb, and Sc
galaxies, respectively, then M$_{\rm sp}$ can be up to $\sim10\%$ of
the mass of the disk. If the corresponding $i$ is around $\sim40\deg$,
$\sim45 \deg$, and $\sim50 \deg$, then M$_{\rm sp}$ is 1\%, 2\% and
3\% of the mass of the disk.  Beyond these values, chaos dominates all
the available phase-space prograde region, destroying the main
periodic and the neighboring quasiperiodic orbits.

All the structural and dynamical parameters of the spiral arms play an
important role in the orbital dynamics. We found however that the
parameter that seems to affect the most the stellar dynamics is the
pitch angle since this presents a wide range of possible observational
values, unlike the case of the mass (or density contrast), and angular
velocity. Within the typical values of the spiral arms angular speed
($\sim 15 - 40 \kmskpc$), obtained from observations and theory, the
chaotic orbital dynamical response does not seem to be extremely
sensitive.

With all the performed simulations we summarize our results, which are
separated according to the two restrictions based on ordered or
chaotic orbital behavior:

\begin {large} \it{Restrictions based on ordered orbital behavior:
The nature of spiral arms, transient or long-lasting.} \end{large}
 
\begin{itemize}

\item [$\bullet$] If the maxima density response (at all radii),
  produced by periodic orbital crowding, support the imposed potential
  of the spiral arms, i.e. if the arms are orbitally self-consistent
  at a first approximation, they are more prone to be long-lasting
  structures; otherwise, the spiral arms would be transient
  structures.
 
\item [$\bullet$] Considering the combination of all parameters
  studied in this work for the spiral arms, we present Table
  \ref{tab:results_order}. The table shows for different types of
  galaxies, the spiral arms persistence based on orbital support.

\item [$\bullet$] All the parameters that characterize the spiral arms
  combined, have to do with the orbital support. From these, due to
  the wide range of values that can take in all galaxies, the one with
  more effect on stellar and gas dynamics seems to be the pitch angle.

\begin{deluxetable}{lcccc}
\tablecolumns{5}
\tabletypesize{\scriptsize}
\tablewidth{0pt}
\tablecaption{Results based on ordered orbital behavior}
\tablehead{{Galactic type} &\multicolumn{3}{c}{Parameter} & }
\startdata
  &$\Omega_{\rm sp}$&Pitch angle &$\mu=M_{\rm sp}/M_{\rm D}$ & Spiral arm persistence\\
  &   ($\kmskpc$)     & ($^{o}$)  & & \\
\hline
 \multirow{4}{*}{Sa} &\multirow{4}{*}{ $20 \lesssim \Omega_{\rm sp}\lesssim 40$}& $\lesssim 10$ &$\lesssim 0.07$ & {Long-lasting} \\
     &&$\lesssim 20$ &$\lesssim 0.02$  &Long-lasting  \\
     &&$\gtrsim 10$&$\gtrsim 0.08$   & {Transient}\\
     &&$\gtrsim 20$&$\gtrsim 0.03$  & Transient\\
 \hline    
  \multirow{4}{*}{Sb} & \multirow{4}{*}{ $15 \lesssim \Omega_{\rm sp}\lesssim 35$}&$\lesssim 15$ & $\lesssim 0.04$ &Long-lasting \\
     &&$\lesssim 25$ &$\lesssim 0.02 $ & Long-lasting  \\
     &&$\gtrsim 15$&$\gtrsim 0.05$&    Transient\\
     &&$\gtrsim 25$&$\gtrsim 0.03$&  Transient\\
 \hline    

 \multirow{4}{*}{Sc} & \multirow{4}{*}{ $10 \lesssim \Omega_{\rm sp}\lesssim 30$}&$\lesssim15$ &$\lesssim 0.04$ & Long-lasting \\
     &&$\lesssim 30$ &$\lesssim 0.01$ &Long-lasting  \\
     &&$\gtrsim15$&$\gtrsim 0.05$   &  Transient\\
     &&$\gtrsim 30$&$\gtrsim 0.02$  & Transient\\

\label{tab:results_order}
\enddata
\end{deluxetable}

\end{itemize}

\begin{large}  \it{Restrictions based on chaotic orbital behavior:
The destruction of spiral arms.} \end{large}

\begin{itemize}

\item [$\bullet$] The main parameters that determine the destruction
  of spiral arms are their pitch angle and mass, both directly related
  to the force amplitude. The destroying effect of their angular speed
  is slight. Table \ref{tab:results_chaos} shows the combination of
  parameters for which chaotic behavior dominates and destroys the
  spiral arms.

\begin{deluxetable}{lcccc}
\tablecolumns{5}
\tabletypesize{\scriptsize}
\tablewidth{0pt}
\tablecaption{Results based on chaotic behavior}
\tablehead{{Galactic type} &\multicolumn{3}{c}{Parameter} & }
\startdata
  
    &$\Omega_{\rm sp}$&Pitch angle &$\mu=M_{\rm sp}/M_{\rm D}$& Chaos predominates\\
  &  ($\kmskpc$)    &  ($^{o}$)  &  & \\
\hline
 \multirow{4}{*}{Sa} &\multirow{4}{*}{ $20 \lesssim \Omega_{\rm sp}\lesssim 40$} &$\lesssim 25$ &$\lesssim 0.09$& {No} \\
     &&$\lesssim 45$ &$\lesssim 0.02$  &No  \\
     &&$\gtrsim 25$&$\gtrsim 0.07$&  Yes\\
     &&$\gtrsim 45$&$\gtrsim 0.01$& Yes\\
 \hline    
  \multirow{4}{*}{Sb} &\multirow{4}{*}{ $15 \lesssim \Omega_{\rm sp}\lesssim 35$} &$\lesssim 35$ &$\lesssim 0.07$ &{No} \\
     &&$\lesssim 45$ &$\lesssim 0.02 $ & No \\
     &&$\gtrsim 35$&$\gtrsim 0.05$  & {Yes}\\
     &&$\gtrsim 45$&$\gtrsim 0.02$  & Yes\\
 \hline    

 \multirow{4}{*}{Sc} &\multirow{4}{*}{ $10 \lesssim \Omega_{\rm sp}\lesssim 30$}&$\lesssim 40$ &$\lesssim 0.05$ & {No} \\
     &&$\lesssim 50$ &$\lesssim 0.03$ & No \\
     &&$\gtrsim 40$&$\gtrsim 0.04$& Yes\\
     &&$\gtrsim 50$&$\gtrsim 0.02$  & Yes\\

\label{tab:results_chaos}
\enddata
\end{deluxetable}

\end{itemize}

  This study searches for the periodic orbits, that are expected to
  be the dynamical backbone of a given system. We search for their
  presence or absence, as a condition for the long-lasting support of
  large-scale structures in a galaxy. In the same manner, when chaos
  dominates the phase space (to such extent that even the main
  periodic orbits are fully destroyed), it is an indication of the
  demolition of large scale structures, such as spiral arms. Although
  it is known that confined chaos (trapped between ordered orbits) is
  able to provide support to structures like spiral arms (Voglis et
  al. 2006), this can only be true as long as chaos does not become
  pervasive.

With all the performed simulations we are able to provide a detailed
set of plausible galactic models (transient or long-lasting), for
normal spiral galaxies, and these idealized galactic models reproduce
astrophysical properties of parameters of observed normal spiral
galaxies, such as the maximum pitch angles observed in spirals. Although one might wonder about the effect of a bar, given the fact
  that bars and spiral arms are formed by disk instabilities, likely,
  even by similar physical processes, the region where a bar grows up
  on a galaxy and the region where spiral arms grow are dominated by
  different physical characteristics (e.g. strong differential
  rotation, mass ratio between spiral arms and the hosting disk,
  structures size and density etc.). In an ongoing work, we include a
  galactic bar potential (combined with the spiral arms). Some
  preliminary results however, show that the presence of a massive bar
  will change dramatically the orbital self-consistency studies and
  new and different restrictions will be likely posed.

\acknowledgments We acknowledge the anonymous referee for an excellent
review that helped to greatly improve this work. We thank PAPIIT
through grant IN114114. APV acknowledges the support of complementary
postdoctoral fellowship of Conacyt at Max Planck Institute for
Extraterrestrial Physics.

\end{document}